\documentclass[
aps,prd,
reprint,
superscriptaddress,% author affiliations via superscript numbers
nofootinbib,
amsmath,amssymb,
floatfix
]{revtex4-2}
\usepackage{upgreek, stix, physics, dsfont, bm} % fonts and symbols

\usepackage{graphicx, xcolor, tikz, tikzscale}
\definecolor{maroon}{HTML}{800020}
\definecolor{darkcyan}{HTML}{008060}
\definecolor{theme1}{rgb}{0.9, 0.36, 0.054} % orange-red
\definecolor{theme2}{rgb}{0.3652, 0.4278, 0.7583} % blue-purple
\definecolor{theme3}{rgb}{0.9451, 0.5939, 0} % yellow-orange
\definecolor{theme4}{rgb}{0.6460, 0.2532, 0.6851} % purple
\definecolor{theme5}{rgb}{0.2858, 0.56, 0.4508} % aqua green
\definecolor{theme6}{rgb}{0.7, 0.336, 0} % brown
\definecolor{theme7}{rgb}{0.4915, 0.3451, 0.8} % cool purple
\definecolor{theme8}{rgb}{0.7179, 0.5687, 0} % dull yellow

\usepackage[colorlinks]{hyperref} % hyperlinking
\hypersetup{colorlinks=true, allcolors=maroon}
\usepackage[capitalise]{cleveref} % avoid \ref, use \cref instead
\crefname{section}{Sec.}{Secs.} % to match PRD crossref style

\newcommand{\Integers}{\ensuremath{\mathds{Z}}}
\newcommand{\Reals}{\ensuremath{\mathds{R}}}

\newcommand{\Ltwo}[1]{\ensuremath{\mathcal{L}^2 \left( #1 \right)}}
\newcommand{\ii}{\ensuremath{\mathrm{i}}}
\newcommand{\ee}{\ensuremath{\mathrm{e}}}
\newcommand{\id}{\ensuremath{\mathds{1}}}
\newcommand{\dbK}{\ensuremath{\mathcal{K}}}
\newcommand{\T}{\intercal} % symbol for the transpose operator
\newcommand{\st}{\ensuremath{\,\middle|\,}}
\DeclareMathOperator{\sgn}{sgn}
\DeclareMathOperator{\spn}{span}

\begin{document}

%---------------------------------------------------------------------------
%   TITLE AND ABSTRACT
%---------------------------------------------------------------------------

\title{Entanglement in quantum field theory via wavelet representations}
%\date{\today}

\author{Daniel~J.~George}
\email[Please direct correspondence to: ]{dan.george@hdr.mq.edu.au}
\affiliation{Department of Physics and Astronomy,
Macquarie University, Sydney, NSW 2109, Australia}
\affiliation{Sydney Quantum Academy, Sydney, NSW 2000, Australia}
\affiliation{ARC Centre of Excellence in Engineered Quantum Systems,
Macquarie University, Sydney, NSW 2109, Australia}

\author{Yuval~R.~Sanders}
\affiliation{Department of Physics and Astronomy, 
Macquarie University, Sydney, NSW 2109, Australia}
\affiliation{ARC Centre of Excellence in Engineered Quantum Systems, 
Macquarie University, Sydney, NSW 2109, Australia}
\affiliation{Centre for Quantum Software and Information,
University of Technology Sydney, Sydney, NSW 2007, Australia}

\author{Mohsen~Bagherimehrab}
\affiliation{Institute for Quantum Science and Technology,
University of Calgary, Calgary, Alberta T2N~1N4, Canada}
\affiliation{Chemical Physics Theory Group, Department of Chemistry, 
University of Toronto, Toronto, Ontario M5G~1Z8, Canada}
\affiliation{Department of Computer Science, 
University of Toronto, Toronto, Ontario M5S~2E4, Canada}

\author{Barry~C.~Sanders}
\affiliation{Institute for Quantum Science and Technology,
University of Calgary, Calgary, Alberta T2N~1N4, Canada}

\author{Gavin~K.~Brennen}
\affiliation{Department of Physics and Astronomy,
Macquarie University, Sydney, NSW 2109, Australia}
\affiliation{ARC Centre of Excellence in Engineered Quantum Systems, 
Macquarie University, Sydney, NSW 2109, Australia}

\begin{abstract}

Quantum field theory (QFT) describes nature using continuous fields,
but physical properties of QFT are usually revealed in terms of 
measurements of observables at a finite resolution. 
We describe a multiscale representation of free scalar bosonic and 
Ising model fermionic QFTs using wavelets. 
Making use of the orthogonality and self-similarity of the wavelet 
basis functions, we demonstrate some well-known relations such as 
scale-dependent subsystem entanglement entropy and renormalization 
of correlations in the ground state. 
We also find some new applications of the wavelet transform as a 
compressed representation of ground states of QFTs which can be 
used to illustrate quantum phase transitions via fidelity overlap 
and holographic entanglement of purification. 

\end{abstract}

\maketitle

%---------------------------------------------------------------------------
%   INTRODUCTION
%---------------------------------------------------------------------------
\section{Introduction}
\label{sec:introduction}

Quantum information has provided new perspectives into quantum field 
theories (QFT), such as using entanglement as a way to characterize 
quantum phases \cite{CC04}, and quantum algorithms for simulating 
scattering cross sections in QFTs that are exponentially faster than
classical algorithms \cite{JLP12}.
Other insights include the study of coarse-graining and renormalization 
from a quantum information perspective \cite{Vid08}, and the harvesting 
of entanglement from vacuum states of QFTs \cite{Rez03,HM20}.
Recently, it was shown that quantum field theories can be 
represented in a way that organizes properties at multiple scales 
using a wavelet functional basis, referred to as a
multiscale representation~\cite{BP13, BRSS15}.
Wavelet-based multiscale representations of QFT have proved 
particularly well suited for studying the holographic principle
\cite{Qi13,LQ16,Lee17,SB16} and renormalization physics \cite{EW16, Alt18}.

Several recent works have demonstrated a connection between wavelets 
and tensor-network- or quantum-circuit-based representations of 
quantum states. 
For example, \textcite{EW16} used a Daubechies wavelet basis to 
analytically construct the tensors in a multiscale entanglement
renormalization ansatz (MERA) description of a ground state of 
massless (critical) fermions on a 1D lattice.
\textcite{HSW+18} showed how to rigorously construct quantum circuits 
that approximate metallic states of massless fermions on 1D and 2D 
lattices based on a discrete wavelet transform using an approximate 
Hilbert pair. 
For quadratic bosonic systems on a lattice, \textcite{WW21} developed
a scale-invariant entanglement renormalization procedure based on
biorthogonal wavelets that disentangles the wavelet output 
at each step. 
Finally, \textcite{WSSW22} found a procedure for constructing MERA-based
quantum circuits that rigorously approximate the continuum correlation
functions for the massless Dirac conformal field theory.

In this paper we derive the wavelet-based multiscale representations of 
two types of QFT: the one-dimensional Ising fermionic QFT and free 
scalar bosonic QFT, both introduced in \cref{sec:background}.
Wavelet-based multiscale representations can be understood as a 
more nuanced form of discretization, in which the continuum Hamiltonian 
is expressed as an infinite number of terms corresponding to ever-finer
length scales.
A minimum length scale is then enforced by the truncation of terms at 
finer length scales.
We demonstrate that a number of established results remain valid
when using these representations, and suggest 
some advantages of such representations for identifying phase transitions.
We present a brief introduction to the relevant aspects of the
wavelet formalism in \cref{sec:wavelets}.

The main results of our paper are contained in \cref{sec:results}.
We show that wavelet-based multiscale representations provide 
natural access to entanglement renormalization physics. 
In \cref{sec:results/correlators} we show numerically that the two-point 
correlators of a coarse-grained QFT decay algebraically in a 
scale-invariant manner at the critical point and with an exponential 
decay with correlation length given by the inverse renormalized mass in 
the massive phase.
In \cref{sec:results/subsystem_entropy} we reproduce using a 
wavelet-based discretization the results of \textcite{CC04} for 
subsystem entanglement in noninteracting bosonic and fermionic QFTs.
\textcite{CC04} model the QFT with a lattice spin system that is treated 
as a discrete approximation to the true continuum theory, justifiable 
by computing a continuum limit (see, e.g., Sec. II in \cite{Boy89}).
We demonstrate that the discretization of these field theories using 
wavelet scale modes reproduces the correct scaling of entanglement 
in both gapped and gapless phases of the theories and we connect the 
phenomenological cutoff length to the scale of our scale modes.

In \cref{sec:results/phases} we consider a multiscale wavelet 
representation of the ground state for a fermionic Ising QFT, and 
show that selection of a subsystem consisting of a small number of 
coarse-grained modes amounts to a form of lossy compression, capturing the
physics of the global pure state up to some error.
We demonstrate the utility of this approach for approximating the 
fidelity overlap between ground states adjacent in some parameter, 
and therefore as a witness for quantum phase transitions, 
where the direct calculation or measurement of fidelity over the global
state may be computationally or experimentally infeasible. 
In \cref{sec:results/holographic_entanglement} we show that 
the entanglement of purification for a reduced quantum state, the
calculation of which quickly becomes unwieldy for large numbers of 
modes, can be well approximated by a coarse-grained state.
This is significant in the context of the work by \textcite{UT18}, 
in which the authors conjecture that the entanglement of purification in
conformal field theories (CFTs) is equal to the minimal-area cross section 
of the entanglement wedge.
Finally, in \cref{sec:discussion,sec:conclusion} we 
summarize our results and conclude with an outlook for further 
applications of our methodology.

%---------------------------------------------------------------------------
%   BACKGROUND
%---------------------------------------------------------------------------
\section{Background}
\label{sec:background}

We focus here on noninteracting one-dimensional fermionic and 
bosonic quantum field theories, due to their mathematical simplicity 
and frequent use as a starting point for perturbative models,
especially in quantum algorithms~\cite{JLP12, BRSS15}. 
They are exactly solvable and therefore allow for direct comparison
of wavelet-based results to known continuum physics.

%---------------------------------------------------------------------------
\subsection{Ising fermionic continuum QFT}
\label{sec:background/fermionic_continuum}

The Hamiltonian density for the free Ising model fermionic quantum field
theory in one dimension is (see Eq. (11a) in \cite{Boy89})
\begin{equation}
  \hat{\mathcal{H}}_\text{f} (x, t) :=
  \frac12 \bigl(
    -\ii \hat{\bm{b}}^T (x,t) \bm{Z} \partial_x \hat{\bm{b}} (x,t)
    + m_0 \hat{\bm{b}}^T (x,t) \bm{Y} \hat{\bm{b}} (x,t)
  \bigr),
\label{eqn:background/fermionic_hamiltonian}
\end{equation}
where $\hat{\bm{b}}(x,t) \equiv 
	\begin{bmatrix} \hat{b}_0 (x,t) \\ \hat{b}_1 (x,t) \end{bmatrix}$
is the spinor of Majorana mode operators at location $x$, $m_0$ is 
the bare mass, and $\bm{Z} = \smqty[\pmat{3}]$ and 
$\bm{Y} = \smqty[\pmat{2}]$ are the usual Pauli matrices.
Note that the spinor components satisfy the equal-time Majorana 
anticommutation relation
\begin{equation}
    \acomm{\hat{b}_\sigma (x)}{\hat{b}_{\sigma'} (x')} =
    2 \delta_{\sigma,\sigma'} \delta(x-x')
\label{eqn:background/majorana_anticommutation}
\end{equation}
for $\sigma,\sigma' \in \{0,1\}$.
The Majorana mode operators have units of inverse square root of length.
In the massless phase, the theory is described by the Ising model CFT 
with central charge $c=1/2$.

%---------------------------------------------------------------------------
\subsection{Free scalar bosonic continuum QFT}
\label{sec:background/bosonic_continuum}

The Hamiltonian density for the free scalar bosonic quantum 
field theory in $d$ spatial dimensions is (see Eq. (11) in \cite{BRSS15})
\begin{equation}
  \hat{\mathcal{H}}_\text{b} (\bm{x}, t) =
  \frac12 \left(
      \hat{\Pi}^2 (\bm{x}, t) +
      \left(\nabla \hat{\Phi} (\bm{x}, t)\right)^2 +
      m_0^2 \hat{\Phi}^2 (\bm{x}, t)
  \right),
\label{eqn:background/bosonic_hamiltonian}
\end{equation}
where the field operator~$\hat{\Phi}(\bm{x}, t)$ and its conjugate 
momentum~$\hat{\Pi}(\bm{x}, t):=\partial_t\hat{\Phi}(\bm{x}, t)$ 
satisfy the canonical equal-time commutation relations
\begin{align}\label{eqn:background/bosonic_commutation}
    &\left[\hat{\Phi}(\bm{x}, t),\hat{\Pi}(\bm{x}', t) \right] =
    \ii \updelta^{(d)}(\bm{x} - \bm{x}') \id,\text{ and}\\
    &\left[\hat{\Phi}(\bm{x}, t),\hat{\Phi}(\bm{x}', t) \right] = 
    \left[\hat{\Pi}(\bm{x}, t),\hat{\Pi}(\bm{x}', t) \right] = 0.
\end{align}
Most of the results below are for the $d=1$ case where the field 
operator is dimensionless. 
In the massless phase, the theory is described by the free bosonic 
CFT with central charge $c=1$.

%---------------------------------------------------------------------------
\subsection{Entanglement entropy scaling}
\label{sec:background/entropy_scaling}

An important physical characterization of QFT is given by 
subsystem entanglement of ground states.
The entanglement entropy of a bipartite pure state 
$\ket{\psi}_{AB}$ is given by the von Neumann 
entropy: $S_A=-\Tr(\rho_A \log \rho_A)$, where the subsystem 
state with support on region $A$ is
$\rho_A=\Tr_B[\ket{\psi}_{AB}\bra{\psi}_{AB}]$. 
For most of the work here, the QFT is assumed to be in one spatial 
dimension over the compact interval $[0,X)$
with specified boundary conditions.
A subsystem $A$ consists of a single subinterval  $[0,X_A)$
of the compact interval $[0,X)$.
The relevant results are given in \textcite{CC04} and \textcite{HLW94}:
\begin{align}
    S_A^\text{critical, periodic} (x) =
    \frac{c}{3} \log \left( \frac{\sin (\pi x)}{\pi \epsilon} \right)
    + \text{constant}
\label{eqn:calabrese-cardy_critical_periodic}
\\
    S_A^\text{critical, open} (x) =
    \frac{c}{6} \log \left( \frac{\sin (\pi x)}{\pi \epsilon} \right)
    + \text{constant}
\label{eqn:calabrese-cardy_critical_open}
\\
    S_A^\text{noncritical} (x) =
    \text{(\# boundary points)} \times \frac{c}{6} \log (m_0 \epsilon)
\label{eqn:calabrese-cardy_noncritical}
\end{align}
which correspond to entropy scaling in the massless~(critical) case, for 
periodic (see Eq. (1) in \cite{CC04}) and open boundary 
conditions (see Eq. (2) in \cite{CC04}), and in the massive~(noncritical) 
case (see Eq. (1) in \cite{CC04}).
Here $\epsilon$ is the ultraviolet cutoff length, 
$c$ the central charge of the relevant CFT, and $m_0$ the mass.
Note also that \cref{eqn:calabrese-cardy_noncritical} is valid only
for the subsystem length $X_A \gg 1/m_0$.

%---------------------------------------------------------------------------
%   WAVELETS
%---------------------------------------------------------------------------
\section{Wavelet-based discretization of quantum field theory}
\label{sec:wavelets}

%---------------------------------------------------------------------------
\subsection{Definition}
\label{sec:wavelets/definition}

Here we use the Daubechies family of wavelets, a family indexed 
by a positive integer $\dbK$ (with $\dbK = 1$ corresponding 
to the well-known Haar wavelet), which has additional beneficial 
properties such as compactness, allowing study of spatially separated 
operators and zero-valued moments.
The Daubechies family of compactly supported wavelets can be defined 
as follows. 
For an alternative introduction, see Sec. 2 in \cite{Bey92}, or for 
a more thorough treatment, see Chap. 7 in \cite{Mal09} or Chap. 5
in \cite{Dau92}.

The wavelet basis is defined in terms of a pair of functions, 
the scale and wavelet functions, here denoted by $s$ 
and $w$ respectively. 
Elsewhere these are sometimes referred to as the father and mother 
wavelets and denoted $\phi$ and $\psi$, respectively.

A function $s \in \Ltwo{\Reals}$ is called a scale function if it 
satisfies the orthonormality condition
\begin{equation}
  \forall \ell \in \Integers: \quad
  \int_\Reals s(x) \, s (x - \ell) \dd{x} = \delta_{0, \ell}
\label{eqn:wavelets/scale_orthonormality}
\end{equation}
and if, for any other function $f \in \Ltwo{\Reals}$,
\begin{align}
  f(x) &\stackrel{\mathrm{a.e.}}{=}
  \lim_{r \to \infty} \sum_{\ell \in \Integers}
  \sqrt{2^r} \, c_\ell^{(r)} \, s \!\left( 2^r x - \ell \right),
\text{ with}\\
  c_\ell^{(r)} &:=
  \sqrt{2^r} \int_\Reals f(x) \, s \!\left( 2^r x - \ell \right) \dd{x},
\label{eqn:wavelets/scale_completeness}
\end{align}
where the symbol $\stackrel{\mathrm{a.e.}}{=}$ means ``equal almost
everywhere,'' which is to say that $f(x)$ is equal to the right-hand 
side for all $x$ except for a measure-zero set. 
For notational convenience, we denote the scale function at
scale (or resolution) $r$ and position $\ell$ by
\begin{equation}
  \forall r, \ell \in \Integers: \;
  s_\ell (x) := s (x - \ell), \;
  s_\ell^{(r)} (x) := \sqrt{2^r} \, s_\ell\!\left(2^r x \right).
\label{eqn:wavelets/scale_notation}
\end{equation}

The scale and wavelet functions at scale $r$ are defined recursively 
as a linear combination of scale functions at scale $r+1$, with 
weights given by the set of scale filter coefficients $\{h_\ell\}$, 
$\ell \in \Integers$: 
\begin{align}
s_\ell^{(r)} (x) &= \sum_{\ell' \in \Integers}
    h_{\ell'} s_{\ell + \ell'}^{(r+1)} (x),
\\
w_\ell^{(r)} (x) &:= \sum_{\ell' \in \Integers}
    (-1)^{\ell'} h_{\Lambda - \ell' - 1} s_{\ell + \ell'}^{(r+1)} (x)
\label{eqn:wavelets/scale_wavelet_definition}
\end{align}
where $\Lambda$ is the number of nonzero filter coefficients such
that $h_\ell = 0$ if $\ell < 0$ or $\ell \geq \Lambda$, 
and a similar notational convention has been adopted for the
wavelet functions as in \cref{eqn:wavelets/scale_notation}.
Specification of these coefficients uniquely determines the 
wavelet basis.

For the Daubechies-$\dbK$ (db$\dbK$) wavelet, the scale filter
coefficients $\{h_\ell\}$ are uniquely determined for $\dbK = 1$,
and up to reflection for any integer $\dbK > 1$, by requiring 
simultaneously that the first $\dbK$ moments vanish:
\begin{equation}
  \int \dd{x} w_0^{(r)} (x) \, x^p = 0,\quad p=0, 1,\ldots,\dbK - 1,
\label{eqn:wavelets/wavelet_moments}
\end{equation}
and that the number of nonzero coefficients, $\Lambda$, is minimized. 
It turns out that this occurs for $\Lambda = 2\dbK$ filter
coefficients. 
Additionally, it can be shown that the scale and wavelet functions at 
scale $0$ are supported on the interval $[0, 2\dbK - 1]$, and that 
the first differentiable scale function is the scale function of the
db$3$ wavelet (see page 239 in \cite{Dau92}), hence its predominant use in this 
paper.

In addition to \cref{eqn:wavelets/scale_orthonormality}, the scale and 
wavelet functions further obey the orthonormality properties
\begin{align}
\forall \ell, \ell', r, r' \in \Integers
  \text{ s.t. } r' \geq r: \;
\int_\Reals s_\ell^{(r)} (x) \, w_{\ell'}^{(r')} (x) \dd{x} = 0,
\\
\forall \ell, \ell', r, r' \in \Integers:
\int_\Reals w_\ell^{(r)} (x) \, w_{\ell'}^{(r')} (x)
  \dd{x} = \delta_{\ell, \ell'} \delta_{r, r'}.
\label{eqn:wavelets/wavelet_orthonormality}
\end{align}

The fixed-resolution subspace $\mathcal{S}_r$ at 
resolution~$r \in \Integers$, along with the associated wavelet 
subspace $\mathcal{W}_r$, can be defined in terms of scale and wavelet
functions at resolution $r$:
\begin{equation}
\mathcal{S}_r := \spn \left\{
    s_\ell^{(r)} \st \ell \in \Integers
  \right\}, \quad
  \mathcal{W}_r := \spn \left\{
    w_\ell^{(r)} \st \ell \in \Integers
  \right\}
\label{eqn:wavelets/fixed_resolution_subspace}
\end{equation}
where $\mathcal{S}_r, \mathcal{W}_r \subset \Ltwo{\Reals}$.
\Cref{eqn:wavelets/scale_wavelet_definition} implies that 
$\mathcal{S}_{r}, \mathcal{W}_{r} \subset \mathcal{S}_{r+1}$, and
from this and the orthogonality conditions in 
\cref{eqn:wavelets/wavelet_orthonormality}, it follows that the
space $\mathcal{W}_{r}$ is precisely the orthogonal complement 
of $\mathcal{S}_{r}$ in $\mathcal{S}_{r+1}$, and therefore
\begin{equation}
  \mathcal{S}_{r} = \mathcal{S}_{r-1} \oplus \mathcal{W}_{r-1}.
\label{eqn:wavelets/wavelet_subspace_isomorphism}
\end{equation}

The wavelet transform in one dimension at scale $r$ is defined as
the isomorphism
\begin{equation}
  \bm{W}^{(r)} :
  \mathcal{S}_r \to \mathcal{S}_{r-1} \oplus \mathcal{W}_{r-1}
\end{equation}
which functions as the basis transform 
\begin{equation}
  \left\{ s_\ell^{(r)} \right\} \to
  \left\{ s_\ell^{(r-1)} \right\} \cup \left\{ w_\ell^{(r-1)} \right\},
  \quad \ell \in \Integers.
\end{equation}
The $d$-level wavelet transform for $d>1$ is defined by the
recursive application of $\bm{W}^{(r)}$ resulting in
\begin{equation}
  \bm{W}^{(r)}_d :
  \mathcal{S}_r \to
  \mathcal{S}_{r-d} \oplus \mathcal{W}_{r-d} \oplus \cdots
  \oplus \mathcal{W}_{r-1}
\label{eqn:wavelets/d-level_wavelet_subspace_isomorphism}
\end{equation}
which functions as the basis transform
\begin{equation}
\begin{split}
  \left\{ s_\ell^{(r)} \right\} &\to
  \left\{ s_\ell^{(r-d)} \right\} \cup
  \left\{ w_\ell^{(r-d)} \right\} \cup
  \left\{ w_\ell^{(r-d+1)} \right\} \\
  & \qquad \cup \cdots \cup
  \left\{ w_\ell^{(r-1)} \right\},
  \ell \in \Integers.
\end{split}
\end{equation}

In this paper we identify $r=0$ with the coarsest scale modes and
consider scales $0 \le r < n$, and therefore make use of 
the $n$-level wavelet transform acting at scale $n$:
\begin{equation}
  \bm{W}^{(n)}_n :
  \mathcal{S}_n \to
  \mathcal{S}_0 \oplus \mathcal{W}_0 \oplus \cdots
  \oplus \mathcal{W}_{n-1}.
\label{eqn:wavelets/n-level_wavelet_subspace_isomorphism}
\end{equation}

The numeric construction of the single-level and multilevel discrete 
wavelet transform from wavelet coefficients is discussed in 
detail in \citeauthor{BSB+22} (see Appendix A in \cite{BSB+22}).

%---------------------------------------------------------------------------
\subsection{Fixed-resolution and multiresolution representations}
\label{sec:wavelets/fixed_multi_representations}

It is useful to define a fixed-scale representation in terms of the 
action of an idempotent projection operator.
Define for each scale $r$ the projection operator mapping from the 
vector space~$\Ltwo{\Reals}$ to the subspace~$\mathcal{S}_r$ 
defined in \cref{eqn:wavelets/fixed_resolution_subspace}:
\begin{equation}
\operatorname{proj}_r :
\Ltwo{\Reals} \to \mathcal{S}_r :
f \mapsto \sum_{\ell \in \Integers} \braket{s_\ell^{(r)}}{f}
    s_\ell^{(r)}
= \sum_{\ell \in \Integers} c_\ell^{(r)} s_\ell^{(r)}
\label{eqn:wavelets/projection_operator}
\end{equation}
where $\ket{s_\ell^{(r)}}$ are the scale functions 
in \cref{eqn:wavelets/scale_wavelet_definition} 
with~$\ell \in \Integers$ and the inner product corresponds to 
the coefficients $c_\ell^{(r)}$ defined 
in \cref{eqn:wavelets/scale_completeness}.
This projection operator is discussed in more depth by \textcite{Dau92}.

Both fermionic and bosonic Hamiltonians include the action of a 
derivative operator, the projection of which requires some care 
since the derivative is not strictly in~$\Ltwo{\Reals}$.
Specifically, the $\alpha$-order derivative
$\dv[\alpha]{}{x}$ acts only upon a proper vector (but not 
Hilbert) subspace of $\Ltwo{\Reals}$ corresponding to the set of 
functions whose derivatives up to $\alpha$ order also belong to
$\Ltwo{\Reals}$.
The projected derivative operator is
\begin{equation}
    \eval{\dv[\alpha]{}{x}}_r :=
    \operatorname{proj}_r \circ
    \dv[\alpha]{}{x} \circ \operatorname{proj}_r,
\label{eqn:wavelets/projected_derivative_operator}
\end{equation}
subject to the requirement that $f$ is a square-integrable function 
with $\alpha$ continuous and square-integrable derivatives. 
Wavelet analysis of these Hamiltonians is therefore restricted to 
scale functions with the requisite properties.
The $\alpha$-order derivative of an arbitrary $\alpha$-order 
differentiable function~$f$ is then
\begin{equation}
\eval{\dv[\alpha]{f}{x}}_r =
    \sum_{\ell,\ell'} \braket{s_\ell^{(r)}}{f} 
    \mel{s_{\ell'}^{(r)}}{\dv[\alpha]{}{x}}{s_\ell^{(r)}}
    s_{\ell'}^{(r)}
\label{eqn:wavelets/projected_a-order_derivative_operator}
\end{equation}
and the action of $\eval{\dv[\alpha]{}{x}}_r$ is entirely
determined by the coefficients
\begin{align}
    \mel{s_{\ell'}^{(r)}}{\dv[\alpha]{}{x}}{s_\ell^{(r)}}
    &=
    \int \dd{x} s_{\ell'}^{(r)} (x) \dv[\alpha]{}{x} s_\ell^{(r)} (x) \\ 
    &=
    2^{\alpha r} \int \dd{x} s_{\ell'-\ell}(x) \dv[\alpha]{}{x} s(x) \\
    &=:
    2^{\alpha r} \Delta^{(\alpha)}_{\ell'-\ell}
    \label{eqn:wavelets/derivative_overlap_coefficients}
\end{align}
henceforth referred to as the derivative overlap coefficients.

Beylkin~\cite{Bey92} showed how to compute the derivative overlap 
coefficients for any scale functions specified by filter 
coefficients $h_\ell, \; \ell \in \{0,\ldots,2\dbK-1\}$ (all 
other coefficients are set to zero) by first defining the 
autocorrelation coefficients (see Eq. (3.19) in \cite{Bey92})
\begin{equation}
    a_\ell := 2 \sum_{\ell'} h_\ell h_{\ell'}.
\end{equation}
Subject to a condition on the number of vanishing moments of the 
resulting wavelet function, Beylkin proved (see Eqs. (4.3)-(4.4) 
in \cite{Bey92}) that the derivative overlap coefficients 
$\Delta^{(\alpha)}_\ell$ constitute the unique solution to the system 
of equations
\begin{align}
\begin{split}
\Delta^{(\alpha)}_\ell
&= 2^\alpha \Delta^{(\alpha)}_{2\ell} +
\frac12 \sum_{k=1}^{\dbK} a_{2k - 1}
\\
& \qquad \times \left( \Delta^{(\alpha)}_{2\ell - 2k + 1} + 
    \Delta^{(\alpha)}_{2\ell + 2k - 1} \right);
\end{split}
\\
\sum_{\ell=-2\dbK+2}^{2\dbK-2} \ell^\alpha \Delta^{(\alpha)}_\ell
&= (-1)^\alpha \alpha!
\label{eqn:wavelets/beylkin_system}
\end{align}
and for a Daubechies wavelet it can be shown
that (see Eqs. (3.51)-(3.52) in \cite{Bey92})
\begin{equation}
\begin{split}
a_{2n-1} &= \frac{ (-1)^{n-1} }
    { (\dbK - n)! (\dbK + n - 1)! (2n-1) }
\\
& \qquad \times \left( \frac{(2\dbK - 1)!}{4^{\dbK - 1}} (\dbK - 1)!
    \right)^2.
\end{split}
\end{equation}
Note that these autocorrelation coefficients are rational, from which 
it follows that the derivative overlap 
coefficients~$\Delta^{(\alpha)}_\ell$ are also rational.

%---------------------------------------------------------------------------
\subsection{One-dimensional Ising fermionic QFT}
\label{sec:wavelets/fixed_scale_fermionic}

A wavelet-based multiscale representation of a continuum Hamiltonian 
over a length interval $[0,X)$ consists of a (countably) infinite sum 
of terms.
Initial terms correspond to the scale modes at the coarsest length scale
(scale $r=0$, length of order $2^{-r}=1$), and progress to wavelet modes 
at finer and finer length scales (down to scale $r=\infty$).
A minimum scale $n$ (length of order $2^{-n}$) is applied by truncating 
terms corresponding to scales $r>n$, and is equivalent to projecting the
Hamiltonian onto the scale subspace~$\mathcal{S}_n$.
The number of modes~$V$ in the system is then
\begin{equation}
    V := 2^n X.
\end{equation}
Equivalently, the Hamiltonian of a system can be directly expressed 
in terms of $V$ scale modes at scale $n$ and the multiscale representation
from scales $0$ to $n$ is then recovered via the application of the 
$n$-level wavelet transform.
This latter process is superficially similar to discretization, and it
is often easier to use the language of discretization 
(e.g. ``wavelet-discretized modes''); however it should be remembered
that the underlying concepts are distinct.

The scale Majorana modes spanning the scale subspace~$\mathcal{S}_r$ 
are defined in terms of the continuum modes like so:
\begin{equation}
    \hat{\bm{b}}_\ell^{(r; \text{s})} :=
    \begin{bmatrix}
    \hat{b}_{\ell, 0}^{(r; \text{s})} \\ \hat{b}_{\ell, 1}^{(r; \text{s})}
    \end{bmatrix},
    \quad 
    \hat{b}_{\ell,\sigma}^{(r; \text{s})} :=
    \int_0^X \dd{x} s_\ell^{(r)} (x) \hat{b}_\sigma (x).
\label{eqn:wavelets/scale_majorana_modes}
\end{equation}
In the fermionic case we consider antiperiodic boundary conditions, 
which correspond to the even parity sector.
The discretized Majorana modes $\hat{b}^{(r; \text{s})}_{\ell,\sigma}$ satisfy
anticommutation relations analogous to those 
of \cref{eqn:background/majorana_anticommutation}
but with the Dirac delta replaced with a Kronecker delta.
Also note that our restriction to a finite-sized subspace of 
$\Ltwo{\Reals}$ ensures that only finitely many of these discretized
Majorana operators are nonzero.

The wavelet Majorana modes are defined analogously by
\begin{equation}
    \hat{b}_{\ell,\sigma}^{(r; \text{w})} :=
    \int_0^X \dd{x} w_\ell^{(r)} (x) \hat{b}_\sigma (x),
\label{eqn:wavelets/wavelet_majorana_modes}
\end{equation}
and complement the scale-discretized Majorana 
modes~$\hat{b}_{\ell,\sigma}^{(r; \text{s})}$.

To project the Hamiltonian from \cref{eqn:background/fermionic_hamiltonian} 
to the subspace~$\mathcal{S}_n$ defined in 
\cref{eqn:wavelets/fixed_resolution_subspace} corresponding to the 
coarse-graining scale~$n$, conjugate $\hat{\mathcal{H}}_\text{f}$ 
with the canonical projection operator $\operatorname{proj}(n)$ 
from \cref{eqn:wavelets/projection_operator}.
This results in the discrete Hamiltonian
\begin{equation}
\begin{split}
\hat{H}^{(n)}_\text{f} &=
- \frac{\ii}{2}\sum_{\ell,\ell' \in \Integers}
    2^n \Delta^{(1)}_{\ell'}
    \hat{\bm{b}}_\ell^{(n;s)}{}^\T \bm{Z}
    \hat{\bm{b}}_{\ell+\ell'}^{(n;s)}
\\
& \qquad + \frac12 m_0 \sum_{\ell \in \Integers}
    \hat{\bm{b}}_\ell^{(n;s)}{}^\T  \bm{Y} \hat{\bm{b}}_\ell^{(n;s)},
\end{split}
\end{equation}
where the symbol $\Delta^{(\alpha)}_\ell$ refers to
the $\ell$th derivative overlap coefficient of order $\alpha$
as defined in \cref{eqn:wavelets/derivative_overlap_coefficients}, 
and $m_0$ is the bare (unrenormalized) mass at scale $n$ and should 
be chosen accordingly.

The quadratic structure of the Hamiltonian can be made explicit by
reexpressing $\hat{H}^{(n)}_\text{f}$ as
\begin{align}
\hat{H}^{(n)}_\text{f} 
&= -\frac{\ii}{2} \sum_{\substack{
    \ell,\ell' \in \Integers\\
    \sigma,\sigma' \in \{0,1\}
    }}
Q_{\ell,\sigma; \ell',\sigma'}^{(n)}
    \hat{b}_{\ell,\sigma}^{(n;s)}  \hat{b}_{\ell',\sigma'}^{(n;s)},
\label{eqn:results/projected_fermionic_hamiltonian}
\\
Q_{\ell,\sigma; \ell',\sigma'}^{(n)} 
&:=
    (-1)^{\sigma} 2^n \Delta^{(1)}_{\ell' - \ell} \delta_{\sigma,\sigma'}
    +m_0 \delta_{\ell,\ell'} (\sigma' - \sigma).
\end{align}
Note that 
$Q_{\ell',\sigma'; \ell,\sigma}^{(n)} =
    - Q_{\ell,\sigma; \ell',\sigma'}^{(n)}$
because $\Delta^{(1)}_{-\ell} = -\Delta^{(1)}_\ell$.

The coefficients $Q_{\ell,\sigma; \ell',\sigma'}^{(n)}$ are 
effectively the entries of a $2V \times 2V$ matrix $\bm{Q}^{(n)}$ 
acting on a vector space with basis vectors indexed by $(\ell,\sigma)$. 
The matrix $\bm{Q}^{(n)}$ is real and antisymmetric and hence has 
pure imaginary 
eigenvalues~$\pm \ii \omega_k^{(n)}, k \in \{\frac12, 
\frac{3}{2}, \ldots, V - \frac12\}$ 
where $\omega_k^{(n)} \in \Reals$.
To find these eigenvalues, observe that due to the applied 
antiperiodic boundary conditions, the submatrix consisting of 
entries~$Q_{\ell, 0; \ell', 0}^{(n)}$ is the antiperiodic analog
of a circulant matrix, with a negative sign applied to the entries 
below the main diagonal.
Hence it can be diagonalized using a half-integer-indexed discrete 
Fourier transform, resulting in submatrix eigenvalues
\begin{equation}
  \lambda_k = \sum_{\ell = 0}^{V - 1} 2^n
  \Delta_{\ell}^{(1)} \ee^{2 \ii \pi k \ell / V}
  = \ii \sum_{\ell = 1}^{2\dbK - 2} 2^{n+1}
  \Delta_{\ell}^{(1)} \sin (\frac{2 \pi k \ell}{V})
\end{equation}
for $k \in \{\frac12, \frac{3}{2}, \ldots, V - \frac12 \}$, 
where the second equality is obtained by using the antisymmetry and
periodicity of the derivative overlap coefficients $\Delta_\ell^{(1)}$ 
and noting that $\Delta_0^{(1)} = 0$.
Similarly the eigenvalues of the submatrix $Q_{\ell, 1; \ell', 1}$ 
are $-\lambda_k$. 
Therefore $\bm{Q}^{(k)}$ is unitarily equivalent to the direct 
sum of $V$ $2\!\times\!2$ matrices:
\begin{equation}
  \bm{Q}^{(n)} \sim \bigoplus_{k = \frac12}^{V - \frac12}
  \begin{bmatrix}
    \lambda_k & -m_0 \\
    m_0 & -\lambda_k
  \end{bmatrix}.
\end{equation}
The eigenvalues of each $2\!\times\!2$ matrix are then
$\pm \ii \omega_k := \pm \sqrt{-m_0^2 + \lambda_k^2}$
and so
\begin{equation}
  \omega_k^{(n)} = \sqrt{m_0^2 + \left(
    \sum_{\ell=1}^{2\dbK-2}
    2^{n+1}\Delta_{\ell}^{(1)} \sin (\frac{2\pi k \ell}{V})
  \right)^2}.
\end{equation}
In the continuum limit as~$n\to \infty$,
\begin{align}
  \omega_k^{(n)} &\to \sqrt{m_0^2 + \left(
    \frac{4 \pi k}{X} \sum_{\ell}
    \ell \Delta_{\ell}^{(1)}
  \right)^2} \nonumber
  \\
  &= \sqrt{m_0^2 + \left(\frac{2 \pi k}{X}\right)^2},
\label{eqn:wavelets/fermionic_continuum_spectrum}
\end{align}
where the first simplification uses the small-angle approximation 
for sine and the second the antisymmetry of the derivative 
overlap coefficients $\Delta_\ell^{(1)}$ together
with \cref{eqn:wavelets/beylkin_system}.
This recovers the well-known dispersion relation for the continuum 
Ising model.

The projected Hamiltonian 
in \cref{eqn:results/projected_fermionic_hamiltonian} is quadratic 
in the fermionic operators. 
Therefore, its ground state is a Gaussian state.
Any Gaussian state is fully characterized by its covariance matrix. 
For a fermionic system, this matrix is obtained from the two-point 
correlations of the system's fermionic operators~\cite{ECP10}.

The entries of the covariance matrix $\bm{\Gamma}^{(n)}$ for 
the ground state of the projected Hamiltonian are defined so that 
\begin{equation}
\Gamma^{(n)}_{\ell,\sigma; \ell',\sigma'}:=
    \ii \expval{\left[\hat{b}_{\ell,\sigma}^{(n;s)}\,,
    \hat{b}_{\ell',\sigma'}^{(n;s)}\right]}
\end{equation}
where~$\expval{\cdot}:=\Tr(\cdot \rho)$. $\bm{\Gamma}^{(n)}$ can be 
calculated by first observing that entries of the covariance 
matrix~$\bar{\bm{\Gamma}}$ for the uncoupled Majorana operators 
$\hat{\bar{\bm{b}}}$ are 
\begin{equation}
\bar{\Gamma}_{\ell,\sigma;\ell',\sigma'} =
    \ii \expval{\left[\hat{\bar{b}}_{\ell,\sigma}\,,
    \hat{\bar{b}}_{\ell',\sigma'}\right]}=
    \updelta_{\ell,\ell'}(\sigma-\sigma').
\end{equation}
which follows from the definition of the ground state and from the 
anticommutation relations for Majorana fermions.

Let~$\bm{O}^{(n)}$ be the orthogonal transformation matrix that
uncouples Majorana operators at scale $n$:
$\hat{\bar{\bm{b}}} := \bm{O}^{(n)} \hat{\bm{b}}^{(n;s)}$.
Then
\begin{equation}
\bm{\Gamma}^{(n)} = \bm{O}^{(n)\T} \bm{\bar{\Gamma}} \bm{O}^{(n)}.
\end{equation}
Calculation of $\bm{O}$ is via symplectic diagonalization
\footnote{
In the case that the eigenvalues of $\bm{Q}^{(n)}$ are all 
nonzero, numeric construction of $\bm{O}^{(n)}$ is simple: of the 
eigenvectors of $\bm{Q}^{(n)}$ having eigenvalues $\pm \ii \omega_k$, 
let $\{\bm{w}_k\}$ be all the eigenvectors corresponding to 
eigenvalues of either $+\ii \omega_k$ or $-\ii \omega_k$. 
The rows of $\bm{O}^{(n)}$ are then the normalized real and imaginary
components of $\bm{w}_k$, that is, 
$\Re \bm{w}_k / \lVert \Re \bm{w}_k \rVert$ followed by 
$\Im \bm{w}_k / \lVert \Im \bm{w}_k \lVert$. 
} 
of the coupling matrix $\bm{Q}^{(n)}$ which takes the analogous form:
\begin{equation}
\bm{Q}^{(n)} =
    \bm{O}^{(n)\T}
    \begin{bmatrix} 0 & \bm{\Omega} \\ -\bm{\Omega} & 0 \end{bmatrix}
    \bm{O}^{(n)},
\label{eqn:fermionic_symplectic_diagonalization}
\end{equation}
where~$\bm{\Omega}$ is the diagonal matrix with diagonal 
entries~$\omega_k$.

Our next step is to approximate the fine-scale correlator
$\expval{ \hat{b}_{\ell,\sigma}^{(n; s)}
    \hat{b}_{\ell',\sigma'}^{(n; s)}}$.
As this is an expectation value of anticommuting operators, its value is
zero whenever $\sigma = \sigma'$.
Translational symmetry requires that
$\expval{\hat{b}_{\ell, 0}^{(n; s)} \hat{b}_{\ell', 1}^{(n; s)}} 
    = \expval{\hat{b}_{0,0}^{(n; s)} \hat{b}_{\ell' - \ell,1}^{(n; s)}}$,
and together with anticommutation it further follows that  
$\expval{\hat{b}_{0, 0}^{(n; s)} \hat{b}_{-\ell, 1}^{(n; s)}}
    = -\expval{\hat{b}_{0, 0}^{(n; s)} \hat{b}_{\ell, 1}^{(n; s)}}$.
We derive expressions for the correlators in
\Cref{sec:fermionic_correlations_derivation}.

%---------------------------------------------------------------------------
\subsection{One-dimensional bosonic QFT}
\label{sec:wavelets/fixed_scale_bosonic}

To obtain a multiscale representation of the bosonic continuum field
theory, we again first define the projection of the bosonic field and
conjugate momentum operators onto the scale subspace~$\mathcal{S}_r$. 
Let
\begin{equation}
\label{eqn:wavelets/bosonic_phi_field_operator}
    \hat{\Phi}^{(r)}_{\ell}(t)
    :=\int_{0}^{X} \dd{x} s_\ell^{(r)}(x) \hat{\Phi}(x,t)
\end{equation}
and
\begin{equation}
\label{eqn:wavelets/bosonic_pi_field_operator}
    \hat{\Pi}^{(r)}_{\ell}(t)
    :=\int_{0}^{X} \dd{x} s_\ell^{(r)}(x) \hat{\Pi}(x,t).
\end{equation}
be the canonical position and momentum field operators projected onto 
the scale subspace~$\mathcal{S}_r$.
The scale field and conjugate momentum operators satisfy commutation 
relations analogous to those of \cref{eqn:background/bosonic_commutation} 
but with the Dirac delta replaced with the Kronecker delta.
The wavelet field and conjugate momentum operators are defined
analogously to 
\cref{eqn:wavelets/scale_majorana_modes,eqn:wavelets/wavelet_majorana_modes}.

Projecting the bosonic Hamiltonian $\hat{\mathcal{H}}_\text{b} (x,t)$ 
from \cref{eqn:background/bosonic_hamiltonian} to a scale subspace
$\mathcal{S}_n$ with the projection operator 
from \cref{eqn:wavelets/projection_operator} results in 
\begin{equation}
    \hat{H}^{(n)}_\text{b} := 
    \frac12 \sum_{\ell \in \Integers}
    \hat{\Pi}_{\ell}^{(n;\text{s})}  \hat{\Pi}_{\ell}^{(n;\text{s})}
    + \frac12 \sum_{\ell,\ell' \in \Integers}
    \hat{\Phi}_{\ell}^{(n;\text{s})}
    K^{(n)}_{\ell \ell'}
    \hat{\Phi}_{\ell'}^{(n;\text{s})},
\label{eqn:discrete_bosonic_Hamiltonian}
\end{equation}
where, for notational simplicity, dependence on time has been dropped, 
and 
\begin{equation}
    K^{(n)}_{\ell,\ell'}
    := m_0^2 \, \updelta_{\ell,\ell'} - 4^n
    \left(
      \Delta^{(2)}_{\ell' - \ell}
    + \Delta^{(2)}_{\ell' - (\ell+N)}
    \right)
\end{equation}
corresponding to periodic boundary conditions.
The spectrum of the projected periodic Hamiltonian is then
\begin{equation}
\omega_k^{(n)} = \sqrt{
    m_0^2 - \sum_{\ell=-2\dbK+2}^{2\dbK-2} 4^n \Delta^{(2)}_{\ell} 
    \cos \left( \frac{2 \pi k \ell}{V} \right)}.
\label{eqn:bosonic_eigenvalues}
\end{equation}
In the continuum limit as~$n\to \infty$,
\begin{align}
\omega_k^{(n)} &\to \sqrt{
    m_0^2 - \sum_{\ell} 4^n \Delta^{(2)}_{\ell} \left( 1 - \frac12 
        \left(\frac{2 \pi k \ell}{V} \right)^2 \right)
}
\\
& = \sqrt{ m_0^2 + \left( \frac{2 \pi k}{X} \right)^2 }
\end{align}
using the cosine small-angle
approximation, \cref{eqn:wavelets/beylkin_system} 
and $\sum_\ell \Delta^{(2)}_\ell =0$ (see Eq. (3.35) in \cite{Bey92}).
In the thermodynamic limit $X \to \infty$, $2\pi k/X$
becomes a continuum parameter that is the momentum of the 
continuum theory, recovering the well-known dispersion 
relation for the continuum bosonic field theory.

The covariance matrix for a bosonic state is defined 
as~\cite{ECP10, FOP05}
\begin{equation}
    \Gamma^{(n)}_{\ell,\ell'}:=
    \expval{\left\{\hat{r}^{(n;\text{s})}_\ell,
        \hat{r}^{(n;\text{s})}_{\ell'}\right\}},
 \end{equation}
where~$\hat{\bm{r}}^{(n;\text{s})} := 
    (\hat{\Phi}^{(n;\text{s})}_0,\ldots,\hat{\Phi}^{(n;\text{s})}_{N-1},
    \hat{\Pi}^{(n;\text{s})}_0,\ldots,\hat{\Pi}^{(n;\text{s})}_{N-1})$ 
is the vector of canonical scale operators at scale $n$.
For the ground state of the Hamiltonian 
in \cref{eqn:discrete_bosonic_Hamiltonian},
the covariance matrix is 
simply~$\bm{\Gamma}^{(n)}=\bm{\Gamma}^{(n)}_{\Phi} 
    \oplus \bm{\Gamma}^{(n)}_{\Pi}$ 
where~$\bm{\Gamma}^{(n)}_{\Pi}:= \sqrt{\bm{K}^{(n)}}$ 
and~$\bm{\Gamma}^{(n)}_{\Phi}
    :=(\bm{\Gamma}^{(n)}_{\Pi})^{-1}$~\cite{ECP10}.

%---------------------------------------------------------------------------
%   RESULTS
%---------------------------------------------------------------------------
\section{Results}
\label{sec:results}

%---------------------------------------------------------------------------
\subsection{Renormalization in multiscale correlators}
\label{sec:results/correlators}

As described in \cref{eqn:wavelets/n-level_wavelet_subspace_isomorphism},
the Hilbert space spanned by scale modes at scale $n$ is 
equivalent to the Hilbert space spanned by coarser scale modes at 
scale $0$ completed by wavelet modes from scales $0 \leq r < n$. 
This allows us to express correlations between wavelet modes at 
some scales $r<n$ in terms of a linear combination of finer scale 
modes at scale $n$. 
This constitutes the so-called bulk/boundary correspondence where 
the wavelet modes (and coarse scale modes) comprise the bulk with 
two dimensions indexed by position and scale, while the finer 
scale modes comprise the one-dimensional boundary with one 
position index. 
The general expression for the correlators of wavelet mode operators 
$\hat{A}^{(\text{w})}$ and $\hat{B}^{(\text{w})}$ at scales~$r$ 
and $r'$ and positions~$\ell$ and $\ell'$ in terms of equivalent 
scale mode operators at scale $n\gg r$ is~\cite{SB16}
\begin{equation}
\begin{split}
&\expval{\hat{A}^{(r;\text{w})}_\ell \hat{B}^{(r';\text{w})}_{\ell'}}
= 2^{-n} 2^{(r+r')/2}
\sum_{j=2^{n-r}\ell}^{2^{n-r}(\ell+2\mathcal{K}-1)}
\sum_{j'=2^{n-r'}\ell'}^{2^{n-r'}(\ell'+2\mathcal{K}-1)}
\\
& \qquad \times w_0^{(0)}(2^{r-n}(j-2^{n-r}\ell+1))
\\
& \qquad \times w_0^{(0)}(2^{r'-n}(j'-2^{n-r'}\ell' +1))
\expval{\hat{A}^{(n;\text{s})}_j \hat{B}^{(n;\text{s})}_{j'}}.
\end{split}
\label{eqn:results/general_correlator}
\end{equation}

%---------------------------------------------------------------------------
\subsubsection{Fermionic case}
\label{sec:results/correlators/fermionic}

For a theory with bare mass $m_0$ defined at the scale $n$, the
fine-scale correlators are (see 
\Cref{sec:fermionic_correlations_derivation} for derivation):
\begin{align}
\expval{\hat{b}_{0, 0}^{(n; s)} \hat{b}_{\ell, 1}^{(n; s)}} 
&= \frac{\ii}{V} 
    \sum_{k \in S} \ee^{-2 \ii \theta_k} \ee^{-\ii 2 \pi \ell k / V},
\\
\text{where }\theta_k 
&= \arctan{\frac{-q_k}{m_0 + \sqrt{m_0^2 + q_k^2}}},
\\
\text{and }q_k 
&= 2 \sum_{j = 1}^{2 \dbK - 2} \Delta^{(1)}_j \sin{\frac{2 \pi j k}{V}}.
\end{align}
In the massless case this simplifies to
\begin{align}
\expval{\hat{b}_{0, 0}^{(n; s)} \hat{b}_{\ell, 1}^{(n; s)}}
&= \begin{cases}
\frac{-2}{V \sin{\left( \pi \ell / V \right)}} & \text{$\ell$ odd}, \\
    0 & \text{$\ell$ even}. \end{cases}
\label{eqn:results/fermionic_massless_correlator}
\end{align}
As per \textcite{SB16} the correlator between wavelet modes at 
different scales $r, r' < n$ is
\begin{align}
\expval{\hat{b}_{0, 0}^{(r; \text{w})}
    \hat{b}_{\ell, 1}^{(r'; \text{w})}}
&\approx \sum_{j,j' = 0}^{V-1}
\expval{\hat{b}_{0, 0}^{(n; s)} \hat{b}_{j - j', 1}^{(n; s)}}
f_{j,0}^{n,r} f_{j',\ell}^{n,r'},
\\
f_{j,\ell}^{n,r}
&= \int \dd{x} s_j^{(n)}(x) w_\ell^{(r)}(x),
\label{eqn:results/fermionic_fine-scale_correlator}
\end{align}
with other expressions given by translational symmetry and anticommutation.
Correlations at the same scale $r=r'$ can be approximated in the continuum 
limit for $r$ sufficiently far from the boundary.
Assuming $r \ll n$, let $2^{r-n} \ell \to x$, treating $x$ as 
a continuous variable so that $\delta \ell= 2^{n-r} \dd{x}$, and 
replacing sums by integrals
$\sum_{m=0}^{2^{n-r}(2\dbK-1)}\to 2^{n-r}\int_0^{2\dbK-1} \dd{x}$,
results in:
\begin{equation}
\begin{split}
\expval{\hat{b}_{0, 0}^{(r; \text{w})}
    \hat{b}_{\ell, 1}^{(r; \text{w})}}
&= \ii 2^{-r} \int_0^{2\dbK-1} \dd{x}
\int_{\ell}^{2\dbK-1+\ell} \dd{x'} w_0^{(0)}(x)
\\
& \qquad \times w_0^{(0)}(x'-\ell)
\expval{\hat{b}_0(x)\hat{b}_1(x')}.
\end{split}
\end{equation}

\begin{figure}
\includegraphics[width=\columnwidth]{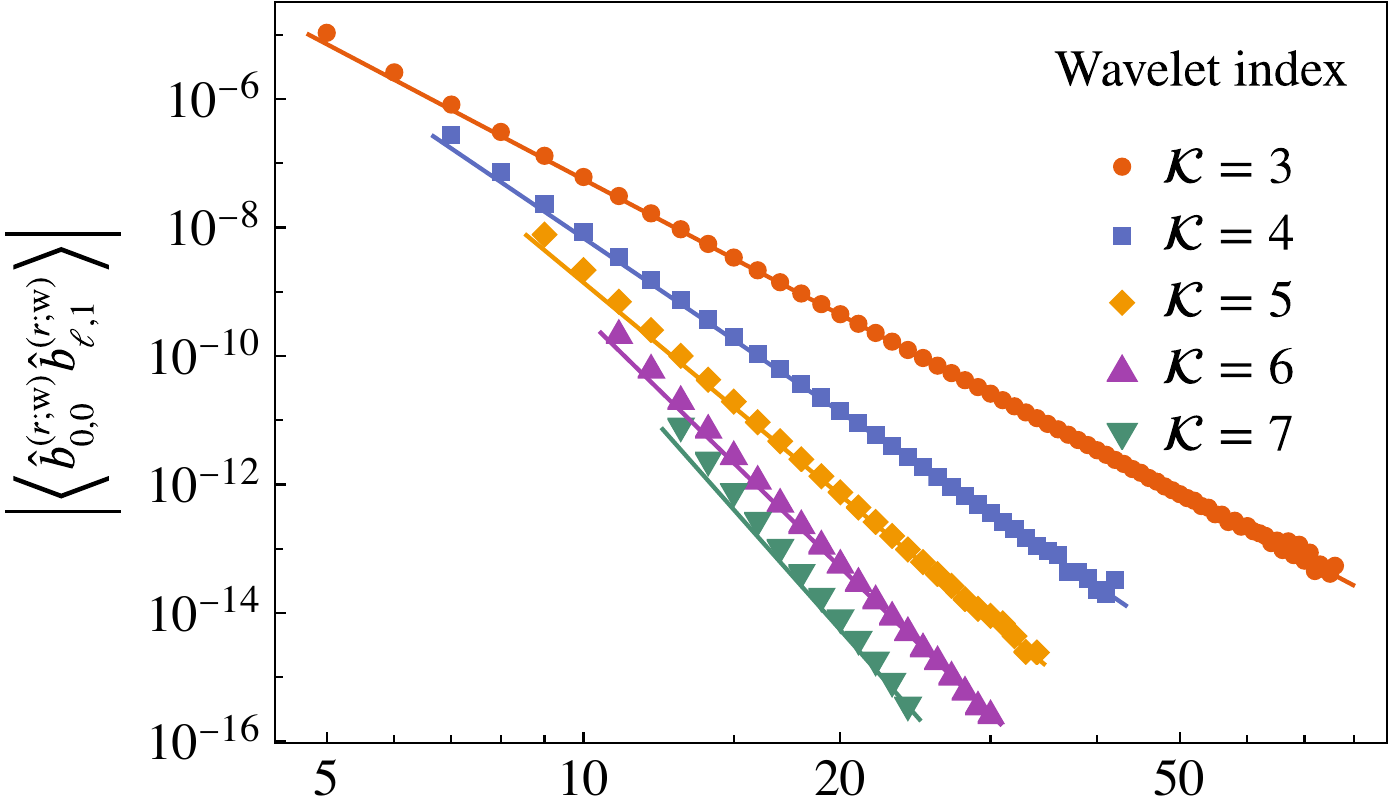}
\includegraphics[width=\columnwidth]{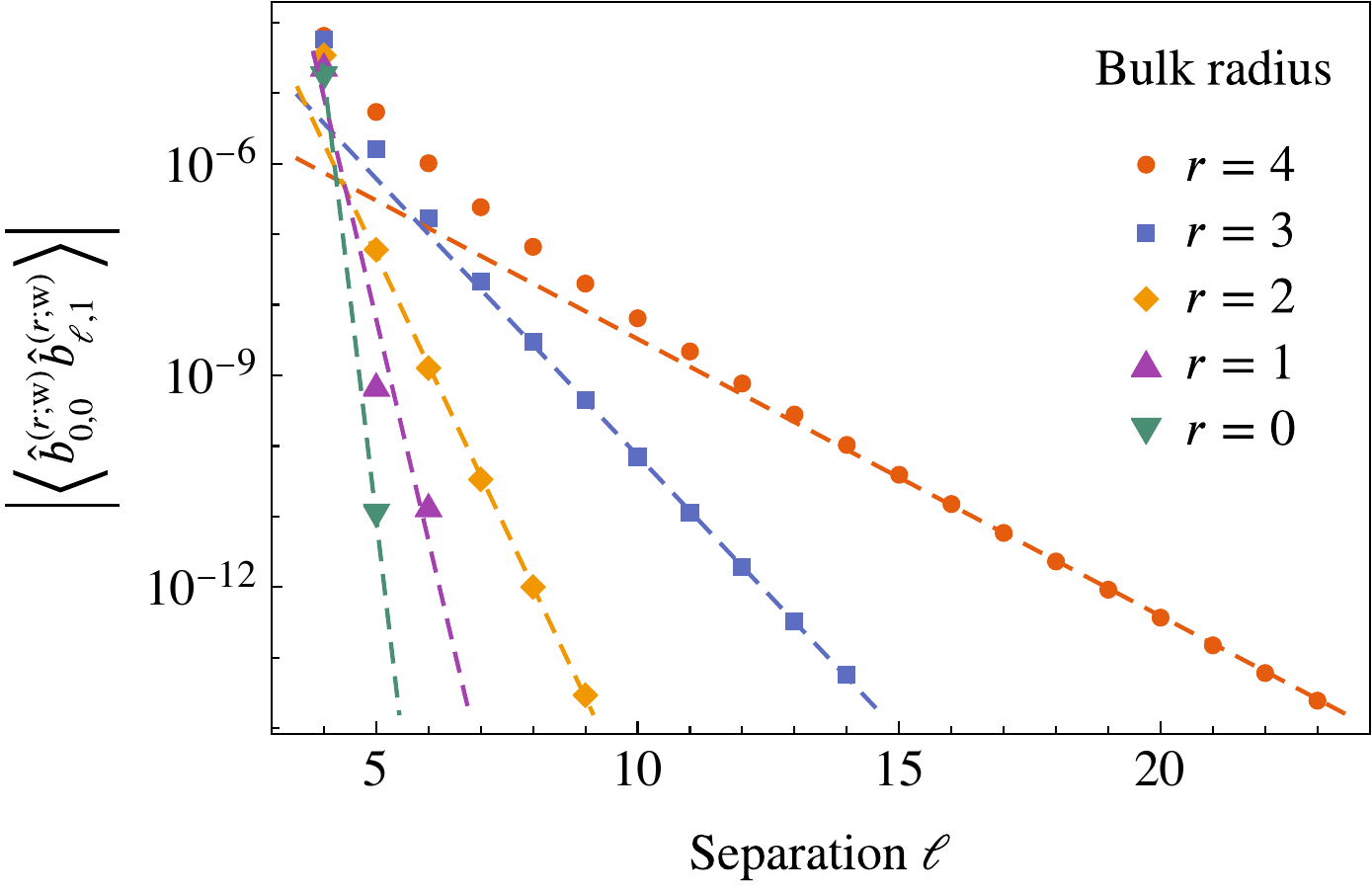}
\caption{\label{fig:results/fermionic_bulk_correlators}
Bulk same-scale wavelet-wavelet correlations 
$\abs{\expval{ \hat{b}^{(r;\text{w})}_{0,0} 
    \hat{b}^{(r;\text{w})}_{\ell,1} }}$ 
at scale $r<n$ as a function of spatial separation~$\ell$ 
for $X=16$, $n=6$, $V=2^n X=1024$.
Top: the critical fermionic Ising model field 
theory~$\hat{H}^{(n)}_\text{f}$, correlations at radius~$r=4$ 
plotted for various Daubechies wavelet indices~$\dbK$.
Solid lines are plots of  
\cref{eqn:results/fermionic_massless_bulk_correlator}. 
Bottom: the massive fermionic field theory with $m_0=0.2$, 
correlations at various radii for $\dbK=3$.
Dotted lines are a joint linear regression assuming $2^{(n-r)}$ dependence.
The correlations fall off exponentially as 
$\ee^{-1.13 \ell \tilde{m}}$, indicating scale-dependent 
mass renormalization, \cref{mrenorm}.
}
\end{figure}

In the massless phase, for large $n$, the continuous correlator can 
be replaced by the discrete correlator 
from \cref{eqn:results/fermionic_massless_correlator}.
Assuming $r \gg 0$ leads to 
$\sin(\pi 2^{-r}(x-x')/V)\approx \pi 2^{-r}(x-x')/V$.
Restrict attention to correlations longer range than the size 
of the wavelet modes, i.e. $\ell>2\mathcal{K}-1$, so that the 
integrals satisfy 
$\int_0^{2\mathcal{K}-1} \dd{x} w_0^{0}(x) w_0^{0}(x-\ell)=0$. 
Define the new variable $x''=x'-\ell$, then
\begin{equation}
\begin{split}
&\expval{\hat{b}_{0, 0}^{(r; \text{w})}
    \hat{b}_{\ell, 1}^{(r; \text{w})}}
= \ii\int \dd{x} \int \dd{x''}
    \frac{w_0^{(0)}(x) w_0^{(0)}(x'')}{\pi ((x-x'')-\ell)}
\\
& \qquad = -\frac{\ii}{\pi \ell}\int \dd{x} \int \dd{x''} w_0^{(0)}(x) w_0^{(0)}(x'')
    \sum_{k=0}^{\infty} \frac{(x-x'')^k}{\ell^{k}}
\\
& \qquad = -\frac{\ii}{\pi}\sum_{k=0}^{\infty} \frac{1}{\ell^{k+1}}
    \sum_{t=0}^k \binom{k}{t} (-1)^{t}\expval{x^t}_w \expval{x^{k-t}}_w.
\end{split}
\label{corrcomp}
\end{equation}
Here $\expval{f(x)}_w\equiv \int f(x) w^0_0(x) \dd{x}$ and for general 
functions must be computed numerically.
However the wavelet moments $\expval{x^a}_w=\int x^a w^0_0(x) \dd{x}$ can 
be computed recursively in closed form (see e.g. \cite{BP13}).
The dominant term in the correlation is determined by the lowest 
nontrivial wavelet moment:
\begin{equation}
    \expval{\hat{b}_{0, 0}^{(k; \text{w})}
        \hat{b}_{\ell, 1}^{(k; \text{w})}} \approx
    \frac{\ii \, (-1)^{\mathcal{K}}}{\pi \ell^{2\mathcal{K}+1}} \times
    \binom{2 \dbK}{\dbK} \expval{ x^\dbK }_w^2.
\label{eqn:results/fermionic_massless_bulk_correlator}
\end{equation}
\Cref{fig:results/fermionic_bulk_correlators} (top) shows this expression 
plotted with direct calculation of the multiscale correlators by 
application of the wavelet transform to the covariance matrix.

For the massive phase, analytic expressions for the correlators are more
difficult to obtain; however, the numerical results, plotted in
\cref{fig:results/fermionic_bulk_correlators} (bottom), demonstrate
exponential falloff with separation with an inverse correlation 
length given by renormalized mass
\begin{equation}
    \tilde{m}=2^{n-r}m_0\label{mrenorm}.
\end{equation}

%---------------------------------------------------------------------------
\subsubsection{Bosonic case}
\label{sec:results/correlators/bosonic}

\begin{figure}
\includegraphics[width=\columnwidth]{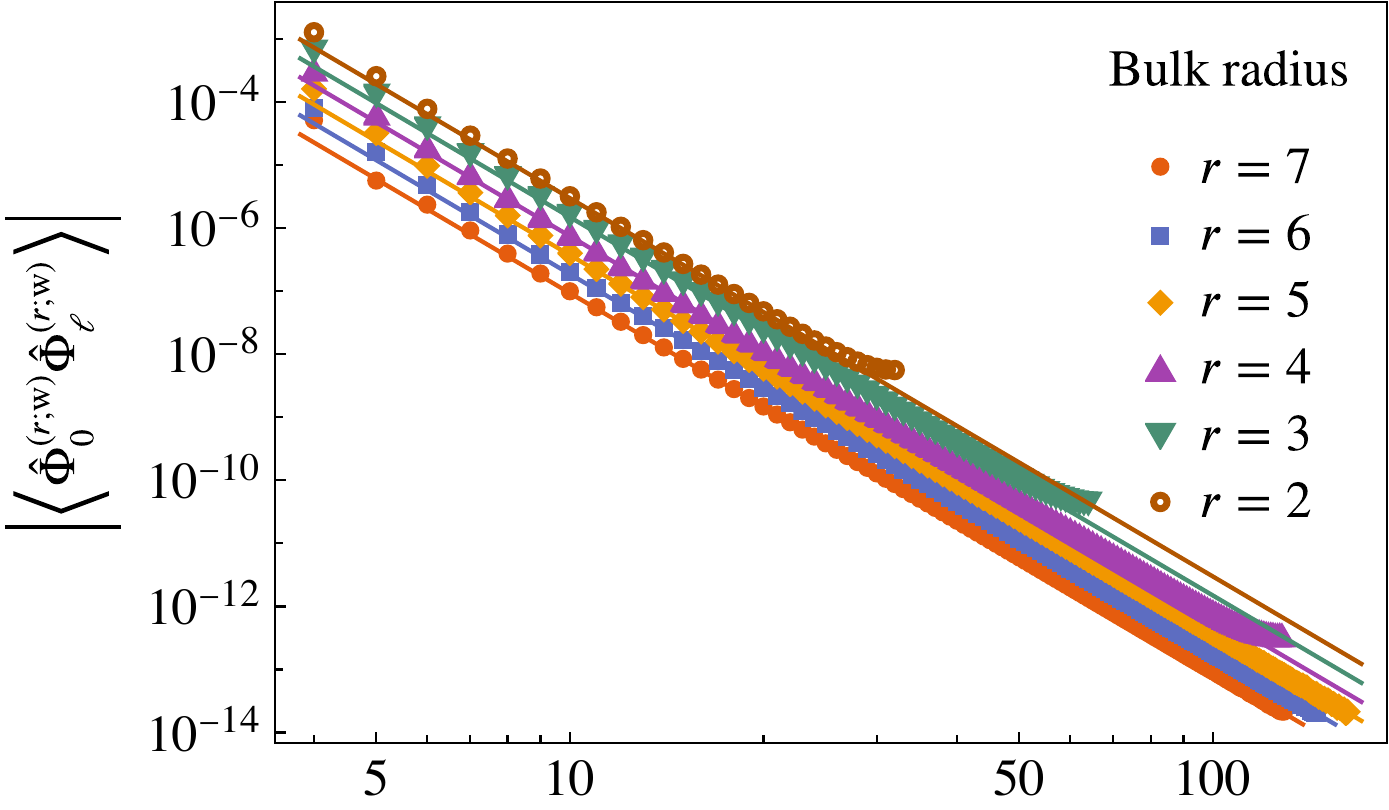}
\includegraphics[width=\columnwidth]{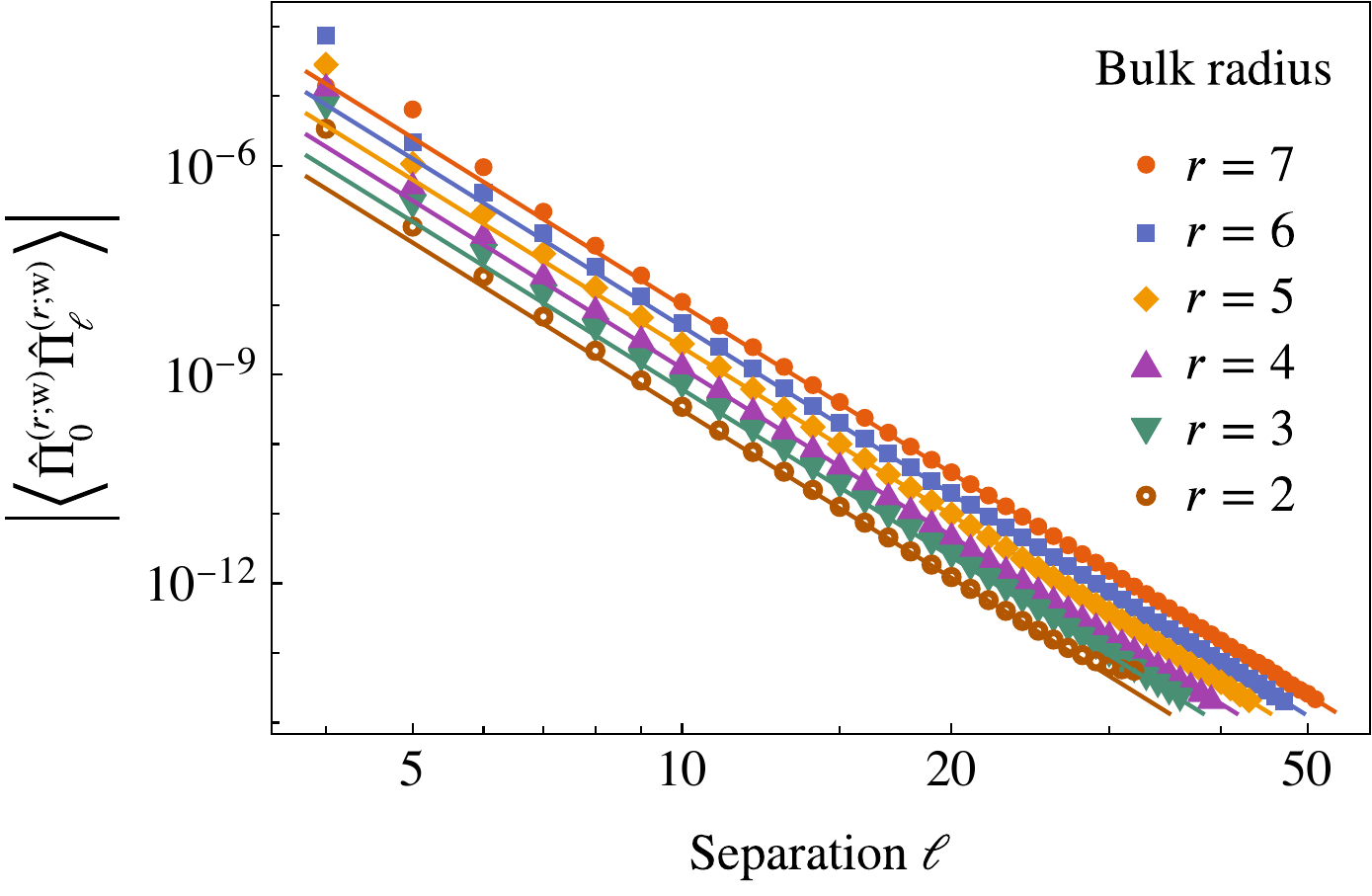}
\caption{\label{fig:results/bosonic_massless_bulk_correlations}
Bulk same-scale wavelet-wavelet correlations for the massless 
bosonic field theory $\hat{H}^{(n)}_\text{b}$ at scale $r<n$ 
as a function of spatial separation~$\ell$ for $X=16$, $n=8$, 
$V=2^n X=4096$ modes with periodic boundary conditions. 
Daubechies $\dbK = 3$ wavelets were used. 
Top: field-field correlations 
$\abs{\expval{ \hat{\Phi}^{(r;\text{w})}_{0}
    \hat{\Phi}^{(r;\text{w})}_{\ell} }}$ with lines 
from \cref{eqn:results/bosonic_massless_correlators_ff}. 
Bottom: momenta-momenta correlations 
$\abs{\expval{ \hat{\Pi}^{(r;\text{w})}_{0}
    \hat{\Pi}^{(r;\text{w})}_{\ell} }}$ with lines 
from \cref{eqn:results/bosonic_massless_correlators_pp}.
}
\end{figure}

\textcite{SB16} (Eqs. B5 and B8 there) show that in the massless phase, the 
same-scale field-field and momenta-momenta correlations for 
separations~$\ell \gg 2\dbK-1$ and $r \ll n$ are
\begin{align}
    \expval{ \hat{\Phi}^{(r;\text{w})}_{0} \hat{\Phi}^{(r;\text{w})}_{\ell} }
    &\approx
    -\frac{ 2^{n-r} }{4 \pi \, \ell^{2 \dbK} \dbK} \times
    \binom{2 \dbK}{\dbK} \expval{ x^\dbK }_w^2,
\label{eqn:results/bosonic_massless_correlators_ff}
    \\
    \expval{ \hat{\Pi}^{(r;\text{w})}_{0} \hat{\Pi}^{(r;\text{w})}_{\ell} }
    &\approx
    \frac{ 2^{r-n} (2 \dbK+1)}{2 \pi \, \ell^{2 \dbK + 2}} \times
    \binom{2 \dbK}{\dbK} \expval{ x^\dbK }_w^2.
\label{eqn:results/bosonic_massless_correlators_pp}
\end{align}
\Cref{fig:results/bosonic_massless_bulk_correlations} shows these 
approximations plotted against direct calculation of the multiscale
correlators from the covariance matrix.

\begin{figure}
\includegraphics[width=\columnwidth]{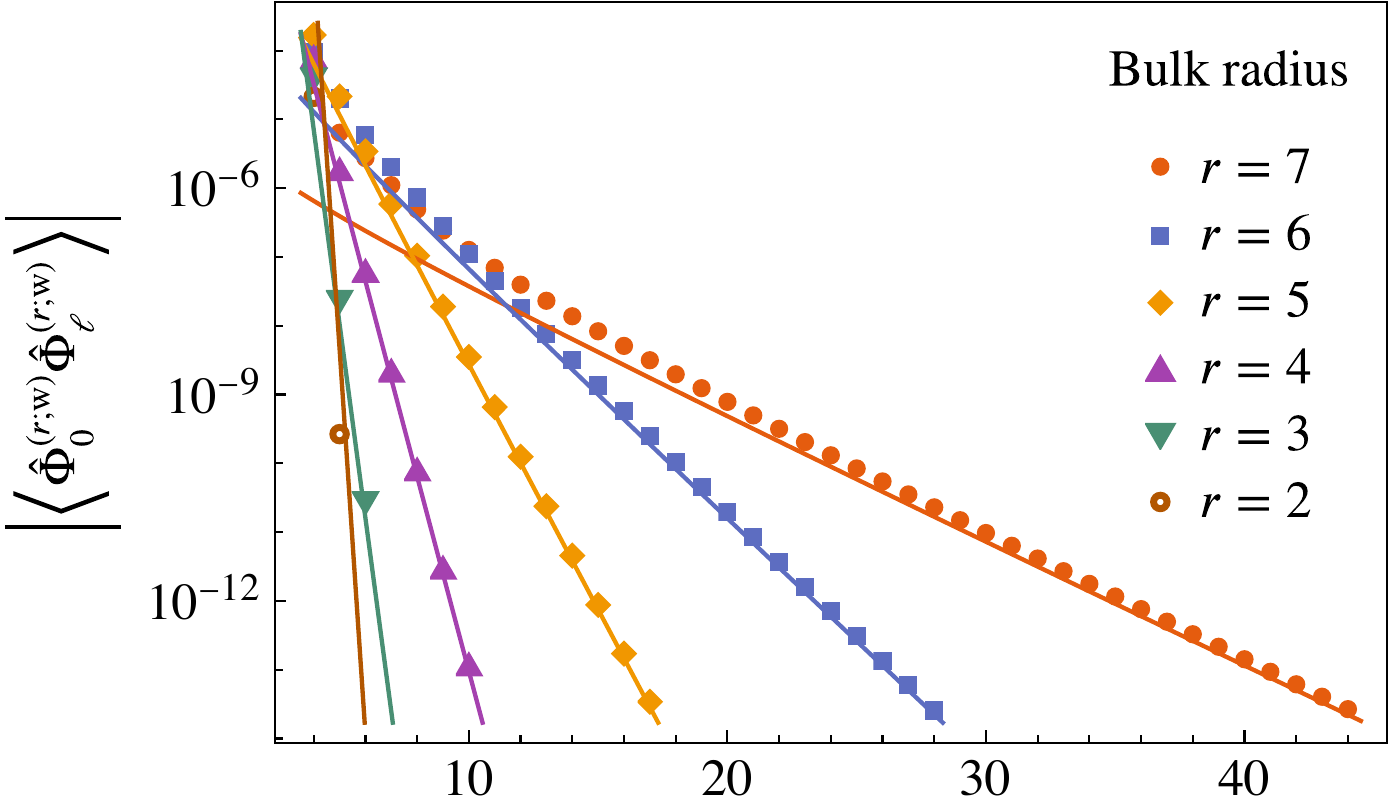}
\includegraphics[width=\columnwidth]{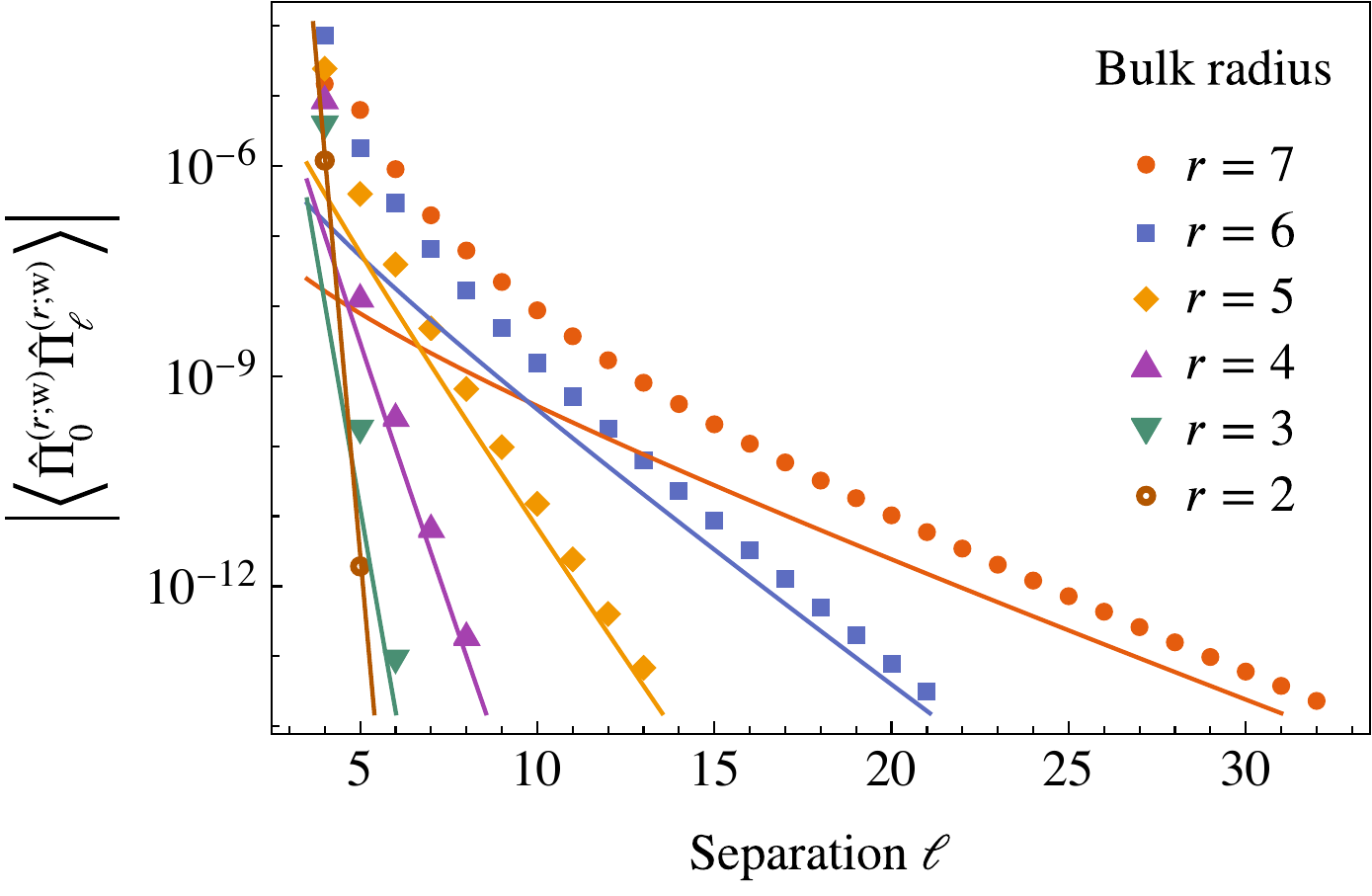}
\caption{\label{fig:results/bosonic_massive_bulk_correlations}
Bulk same-scale wavelet-wavelet correlations for the massive 
bosonic field theory $\hat{H}^{(n)}_b$ at scale $r<n$ as a 
function of spatial separation $\ell$ for $X=16$, $n=8$, 
$V=2^n X=4096$ modes with periodic boundary conditions. 
Mass is $m_0 = 0.2$ and Daubechies $\dbK = 3$ wavelets were used. 
Top: field-field correlations 
$\abs{\expval{ \hat{\Phi}^{(r;\text{w})}_{0}
\hat{\Phi}^{(r;\text{w})}_{\ell} }}$ with lines 
from \cref{eqn:results/bosonic_massive_correlators_ff}. 
Bottom: momenta-momenta correlations 
$\abs{\expval{ \hat{\Pi}^{(r;\text{w})}_{0}
\hat{\Pi}^{(r;\text{w})}_{\ell} }}$ with lines 
from \cref{eqn:results/bosonic_massive_correlators_pp}.
}
\end{figure}

In the massive phase, the bulk correlations are exponentially decaying 
in all directions.
For $\tilde{m} \gg 1$ and separations~$\ell \gg 2\dbK-1$, the 
coarse-grained ($r \ll n$) field-field and momenta-momenta 
correlations are (see Eqs. B15 and B16 in \cite{SB16})
\begin{align}
    \expval{ \hat{\Phi}^{(r;\text{w})}_{0} \hat{\Phi}^{(r;\text{w})}_{\ell} }
    &\approx
    -\frac{ 2^{n-r} e^{-\ell \tilde{m}} }{\sqrt{8 \pi \ell \tilde{m}} }
        \expval{e^{-\tilde{m} x}}_w \expval{e^{\tilde{m} x}}_w
\label{eqn:results/bosonic_massive_correlators_ff},
    \\
    \expval{ \hat{\Pi}^{(r;\text{w})}_{0} \hat{\Pi}^{(r;\text{w})}_{\ell} }
    &\approx
    2^{r-n} e^{-\ell \tilde{m}} \sqrt{\frac{\tilde{m}}{8\pi \ell^3}}
        \expval{ e^{-\tilde{m} x}}_w \expval{e^{\tilde{m} x}}_w.
\label{eqn:results/bosonic_massive_correlators_pp}
\end{align} 
\Cref{fig:results/bosonic_massive_bulk_correlations} shows these 
expressions plotted with numerical calculations of the
multiscale covariance matrix.

\begin{figure}
\includegraphics[width=\columnwidth]{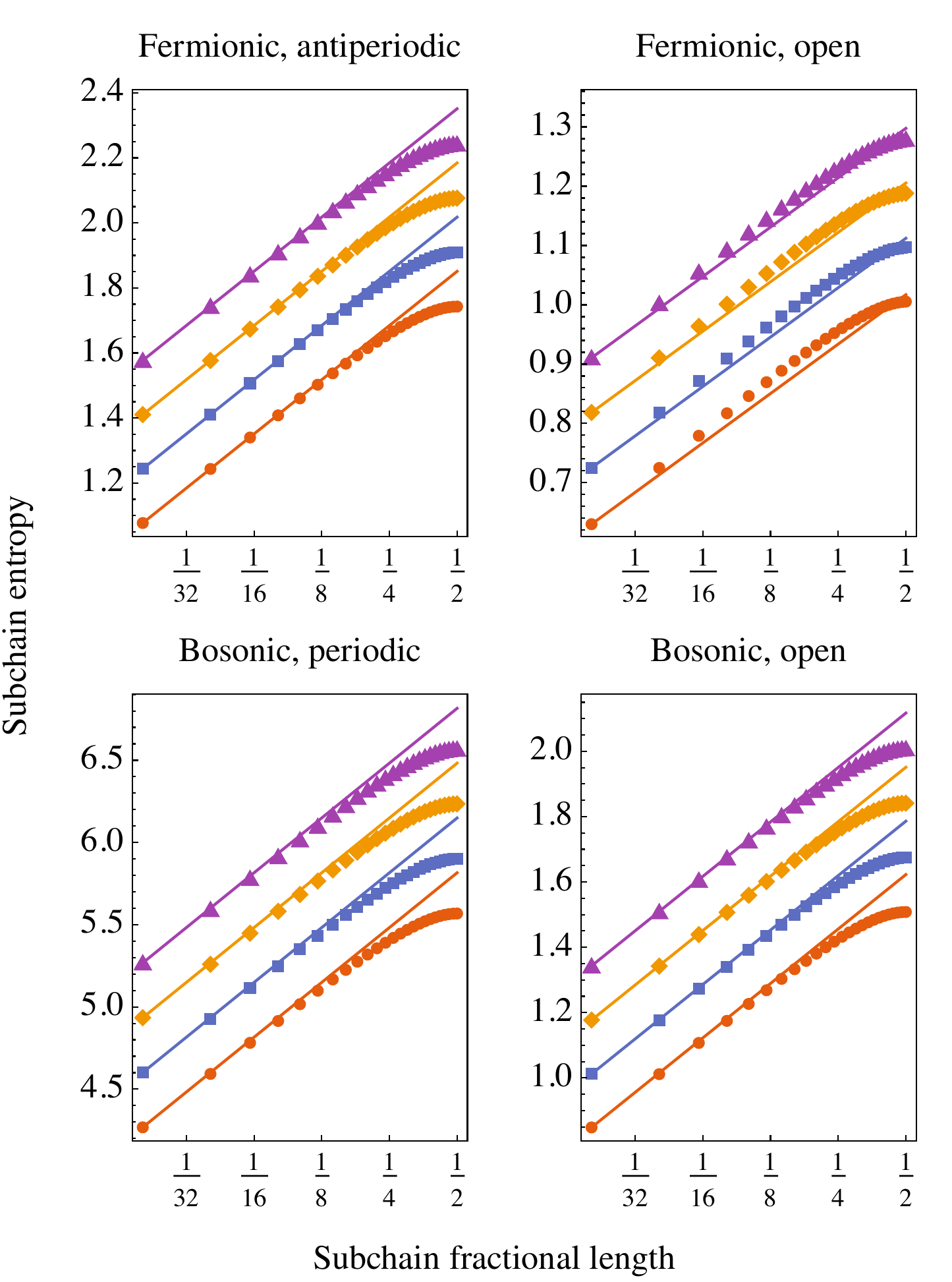}
\caption{\label{fig:results/subsystem_entropy}
Plots of the critical subsystem entropy for the fermionic and 
bosonic theories with (anti)periodic and open boundaries.
Each of the data points correspond to a subinterval 
of the indicated fractional length.
A boundary point of the subinterval always coincides with a 
boundary point of the whole interval.
Each curve depicts the system at a different resolution with number of 
modes $V=500$ (orange circles, bottom), $V=1000$ (blue squares), 
$V=2000$ (yellow diamonds), or $V=4000$ (purple triangles, top). 
Entropy increases with resolution.
In the (anti)periodic case, we encounter a singularity in the zero-mass 
limit and therefore set $m_0 = 10^{-8}$ for the fermionic system and
$m_0 = 10^{-4}$ for the bosonic system.
The solid lines are used to indicate the central charge, and have slope
$1/6$ in the antiperiodic fermionic case, $1/12$ in the open fermionic 
case, $1/3$ in the periodic bosonic case and $1/6$ for the open bosonic 
cases.
These slopes correspond to the expected central charges of $c=1/2$ for 
the Ising model CFT and $c=1$ for the free bosonic CFT in the context 
of \cref{eqn:calabrese-cardy_critical_periodic,%
eqn:calabrese-cardy_critical_open}.
The leftmost point of each line is chosen to match the 
leftmost data point.
}
\end{figure}

%---------------------------------------------------------------------------
\subsection{Entanglement entropy of subsystems in 1D}
\label{sec:results/subsystem_entropy}

%---------------------------------------------------------------------------
\subsubsection{Subsystem specification}

In the wavelet scale basis, as in the canonical basis, the covariance
matrix~$\bm{\Gamma}^A$ of a subsystem~$A$ is obtained by selecting a
subset of the rows and columns of the system's covariance 
matrix~$\bm{\Gamma}$. 
Suppose that the subsystem~$A$ of a system containing~$V$ scale modes 
at scale $n$ over the interval $[0,X)$ is an interval of the form
$[x_\text{min},x_\text{max}) := 
[\ell_\text{min}/2^n, \ell_\text{max}/2^n)$ for integers
$0 \leq \ell_\text{min} < \ell_\text{max} \leq V$.
The subset of modes $s_\ell^{(n)}$ retained belong to the
interval~$[x_\text{min},x_\text{max})$ and correspond 
to~$\ell \in \{\ell_\text{min},\ldots,\ell_\text{max}-1\}$.
Setting~$V_A := \ell_\text{max} - \ell_\text{min}$, 
the resulting covariance matrix~$\bm{\Gamma}^A$ is of size
$V_A \times V_A$.
The total length of the interval~$[x_\text{min},x_\text{max})$, 
or length of the subsystem~$A$, is~$X_A = V_A/2^n$.

In these calculations a subsystem~$A$ with $V_A \leq V/2$ modes 
is selected to be an interval of the form~$[0,X_A)$. 
The subsystem's modes are taken to be less than~$V/2$ as the entropy 
of two subsystems with~$V/2-V_A$ and~$V/2+V_A$ modes are the same.
\Cref{fig:results/subsystem_entropy} shows the entropy plots 
for the massless bosonic and fermionic theories with different 
boundary conditions.

%---------------------------------------------------------------------------
\subsubsection{Fermionic case}

The covariance matrix~$\bm{\Gamma}$ for a fermionic state is real 
and antisymmetric, and 
satisfies~$\bm{\Gamma}^2\geq -\id$ (\cite{MEC+16}, page 2).
Therefore its eigenvalues are all purely imaginary and come 
in positive and negative pairs; i.e., the set of eigenvalues 
is~$\{\pm\ii \sigma_\ell\}$. 
Moreover,~$ \abs{\sigma_\ell} \leq 1$.
Note that~$\{\pm \sigma_\ell\}$ is the set of singular values 
of~$\bm{\Gamma}$ which is equal to the set of 
eigenvalues of~$\ii\bm{\Gamma}$.
The entanglement entropy of an~$N$-mode fermionic Gaussian 
state~$\rho^A$ can then be expressed in terms of the singular 
values~$\{\sigma^A_\ell\}$ of its covariance matrix~$\bm{\Gamma}^A$ 
as (\cite{MEC+16}, page 2)
\begin{align}
\label{eqn:fermionic_entropy}
S\left(\rho^A\right)= 
  \sum_{ \sigma^A_\ell \in [0,1]}
  H\left(\frac{1+\sigma^A_\ell}{2} \right),
\end{align}
where~$H(x):= -x\log_2(x)-(1-x)\log_2(1-x)$ is the binary 
entropy function.

For the massive fermionic theory, we observe that the entanglement 
entropy of a subsystem is constant as a function of the subsystem's 
length in different scales, and this constant increases with the scale~$r$. 
This is expected from the entanglement area law.
For the massless theory, we observe that the functional form of the
Calabrese-Cardy relations are correct as given in
\cref{sec:background/entropy_scaling}, and the central charge is 
correct by means of a line with slope equal to~$c/3$ and~$c/6$,
for periodic and open boundary conditions respectively. 
\Cref{fig:results/subsystem_entropy} shows the entropy plots for 
the massless theory with different boundary conditions.

%---------------------------------------------------------------------------
\subsubsection{Bosonic case}

The covariance matrix~$\bm{\Gamma}$ for a bosonic state is a real 
and positive-definite symmetric matrix, and
satisfies~$\bm{\Gamma}+\ii\bm{\Omega}/2 \geq 0$~(\cite{FOP05}, page 2)
where
\begin{align}
    \bm{\Omega}:= \begin{pmatrix}
    \bm{0} & \id_{N} \\
    -\id_{N} & \bm{0}
    \end{pmatrix}.
\end{align}
Williamson's theorem states that any symmetric and positive-definite 
matrix, such as the covariance matrix~$\bm{\Gamma}$, can be decomposed
as~$\bm{\Gamma} = \bm{S}^T (\bm{\Lambda} \oplus \bm{\Lambda})
\bm{S}$~(\cite{FOP05}, page 18),
where~$\bm{S}$ is a symplectic matrix and~$\bm{\Lambda}$ is a 
diagonal matrix whose spectrum~$\{\lambda_\ell\}$ is equal to 
the set of positive eigenvalues 
of~$\ii \bm{\Omega} \bm{\Gamma}$~(\cite{ECP10}, page 281).
The eigenvalues of~$\bm{\Lambda}$ are called the 
symplectic eigenvalues of the covariance matrix~$\bm{\Gamma}$. 
$\bm{\Gamma}+\ii\bm{\Omega}/2 \geq 0$ further implies 
that~$\lambda_\ell \geq 1/2$.

The entanglement entropy of an~$N$-mode bosonic Gaussian 
state~$\rho^A$ can then be expressed in terms of the symplectic 
eigenvalues of its covariance matrix~$\bm{\Gamma}^A$ 
as~\cite{Dem18}
\begin{align}
S\left(\rho^A\right) = 
 \sum_{\ell=1}^{2N} f\left(\lambda^A_\ell\right),
\label{eqn:bosonic_entropy_wrt_sympEigens}
\end{align}
where
\begin{align}
    f(\lambda):=(\lambda + \frac12) \log_2 (\lambda + \frac12)
    - (\lambda - \frac12) \log_2 (\lambda - \frac12).
\end{align}

Similar to the fermionic case, we again observe that the functional 
form of the Calabrese-Cardy relations are correct as given in
\cref{sec:background/entropy_scaling}. 

%---------------------------------------------------------------------------
\subsubsection{Cutoff scaling behavior}

For a massless bosonic field theory in one spatial dimension 
with periodic boundary conditions the 
subsystem entanglement is related to the ultraviolet cutoff
parameter $\epsilon$ as per \cref{eqn:calabrese-cardy_critical_periodic}
with $c=1$.
We computed the half-chain entropy in a wavelet basis ($\dbK = 3$)
for resolutions $V=2^n X$ with $X=32$ 
and for $n=0,\ldots, 7$ (mass was technically $m_0=10^{-4}$ to avoid
a singularity in the zero-mass limit). 
The numerical results demonstrate linear scaling in
$n$ with a least-squares fit of $S = 6.23 + 0.333 n$, in agreement
with the expected scaling law 
$\epsilon \propto \frac{1}{V} = \frac{1}{2^n X}$.

%---------------------------------------------------------------------------
\subsection{Entanglement entropy in 2D}

\begin{figure}
\includegraphics[width=\columnwidth]{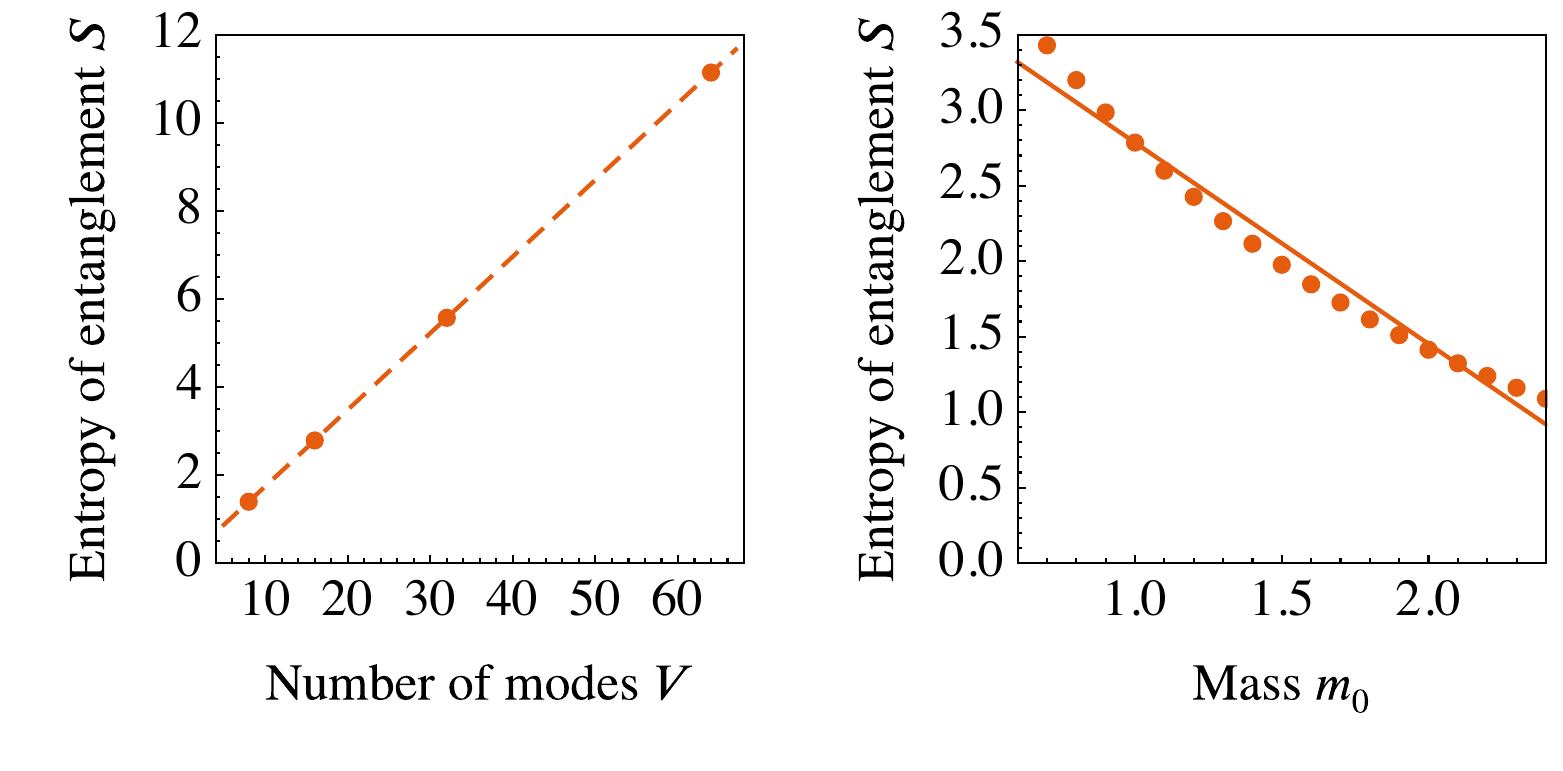}
\caption{\label{fig:results/bosonic_subsystem_entropy_2d}
Entropy of entanglement for a $V\!\times\!\frac{V}{2}$-mode 
subsystem of a $V\!\times\!V$ two-dimensional system as a function of the 
number of modes $V=2^n X$, with $X=8$.
The one-dimensional area of the boundary is $V$.
Left: entropy of entanglement against number of modes for
$n \in \{0,1,2,3\}$, with mass $m_0=1$, demonstrating expected 
linear scaling of $S$ with $V$ 
(dashed line is the least-squares fit corresponding to $\epsilon = 3.88$). 
Right: same as left but as a function of mass with $n=1$.
The solid line is
\cref{eqn:results/entanglement/eescalingtwodim}, with a slope of
$-A_\perp/12 = -4/3$ and the intercept is
$A_\perp/\epsilon = 4.12$ using $\epsilon$ from the fit on the
left (note the values coincide at $m_0=1$ as expected).
While the slope is reasonably accurate over this mass range, it is 
apparent higher order terms contribute.
}
\end{figure}

Since there is no direct analog for the fermionic Ising model in
two dimensions, we consider only bosonic systems here.
The scaling of subsystem entanglement entropy depends on the number
of spatial dimensions. 
For two spatial dimensions, to first order in $A_\perp$, the scaling 
for a free bosonic QFT is~\cite{Her13}
\begin{equation}
    S \approx \frac{A_\perp}{\epsilon} - \frac{1}{12} A_\perp m_0,
\label{eqn:results/entanglement/eescalingtwodim}
\end{equation}
where $A_\perp$ is the area of the one-dimensional subsystem 
boundary and $\epsilon$ is the ultraviolet cutoff of the field theory.
The scaling behavior of the massive bosonic field theory in two 
dimensions in a wavelet basis is demonstrated 
in \cref{fig:results/bosonic_subsystem_entropy_2d}.

%---------------------------------------------------------------------------
\subsection{Discriminating quantum phases via fidelity overlap}
\label{sec:results/phases}

One witness of a quantum phase transition (QPT) is a sudden drop 
in the overlap fidelity between ground states of Hamiltonians 
straddling a critical point \cite{ZP06}.
Specifically, for a Hamiltonian which experiences a QPT as a 
function of one parameter $g$, the witness is
\begin{equation}
F = \abs{\braket{\Psi(g_+)}{\Psi(g_-)}},
\end{equation}
where $g_{\pm}=g\pm\delta/2$, and $\ket{\Psi(g)}$ is the ground 
state of the Hamiltonian with parameter $g$. Here $\delta$ is 
some increment small enough to resolve the change in $F$.

As in \Cref{sec:fermionic_correlations_derivation}, the ground state 
of the fermionic Ising model, where the relevant parameter is $g=m_0$, 
is specified by the condition 
\begin{equation}
\hat{\eta}_k\ket{\Psi}=0,
\end{equation}
where the normal fermionic modes are defined in terms of the 
momenta annihilation and creation 
operators~$\hat{p}_k, \hat{p}_k^{\dagger}$ as
\begin{align}
\hat{\eta}_k &= \cos\theta_k\hat{p}_k +
    \ii\sin\theta_k\hat{p}_{-k}^{\dagger},
\\
\theta_k &= \cos^{-1} \left[
    \frac{-(m_0+\omega_k)}{\sqrt{(m_0+\omega_k)^2+q_k^2}} \right],
\\
q_k &= 2\sum_{\ell=1}^{2\dbK-2}\Delta^{(1)}_{\ell}
    \sin\left( \frac{2\pi k\ell}{V} \right).
\end{align}
Negative momenta modes are defined by positive indexing via $-k\equiv V-k$.
For antiperiodic boundary conditions the set of allowable positive 
momenta $k$ is 
$k \in S_L=\{\frac{1}{2},\frac{3}{2},\ldots,\frac{V-1}{2}\}$. 
Using the inverse Jordan-Wigner transformation and ordering qubits 
in pairs $\{(k,-k)\}_{k\in S_L}$ results in expressions 
$\hat{p}_k = [\prod_{j<k} \bm{Z}_j \bm{Z}_{-j}] \sigma^+_k$ and
$\hat{p}_{-k}^{\dagger} = [\prod_{j<k}
  \bm{Z}_j \bm{Z}_{-j}] \bm{Z}_k \sigma^-_{-k}$,
where $\sigma^{\pm}_k \equiv \frac{1}{2}(\bm{X}_k \pm \ii \bm{Y}_k)$ are
the usual raising and lowering operators acting on mode $k$.
The ground state can then be written as a tensor product of 
entangled qubit pairs
\begin{equation}
    \ket{\Psi} = \bigotimes_{k\in S_L} (
    \cos\theta_k\ket{1}_k\ket{1}_{-k} + 
    \ii\sin\theta_k\ket{0}_{k}\ket{0}_{-k}
    ).
\end{equation}
The fidelity is then
\begin{equation}
    F = \prod_{k\in S_L} \cos(\theta_k(m_+) - \theta_k(m_-)),
\label{eqn:analytic_fidelity_overlap}
\end{equation}
where $\theta_k(m_{\pm})$ is the value $\theta_k$ with 
mass~$m_0 \pm \delta/2$. 
Most of the terms in the product formula for the fidelity are equal 
to one up to numerical precision for any mass, but near $m_0 = 0$ there 
are deviations which are most prominent at the longest wavelength, 
namely $|k|=1/2$.
Approximating the fidelity at criticality by the overlap on this single 
pair of modes and assuming $1/\delta,V\gg 1$ results in
\begin{equation}
    F \approx 1-\frac{\delta^2 V^2}{8\pi^2}, \quad m_0 = 0,
\end{equation}
where the derivative coefficients have been summed as 
in \cref{eqn:wavelets/beylkin_system}.
By contrast, away from criticality, letting $\delta=1/V$,
\begin{equation}
    F \approx 1-\frac{\pi^2\delta^2}{8m_0^4 V^2}, \quad \abs{m_0} \gg 1/V,
\end{equation}
which quickly approaches $1$.

Rather than computing the fidelity overlap of the global ground states, 
an approximation can be obtained in a multiresolution wavelet basis 
by computing the fidelity overlap between reduced states of a 
few coarse modes.
This reduced state effectively acts as a compressed representation 
of the global state. 
Specifically, the fidelity between reduced states is given by:
\begin{equation}
F(\rho_s(g_+),\rho_s(g_-))
= \Tr[\sqrt{ \sqrt{\rho_s(g_+)} \rho_s(g_-) \sqrt{\rho_s(g_+)} }],
\end{equation}
where $\rho_s(g)$ is the reduced state on subsystem $s$ of the global 
pure state $\ket{\Psi(g)}$.
For the fermionic Ising model field theory, the fidelity between two 
mixed fermionic Gaussian states with covariance matrices 
$\Gamma_s(g_+),\Gamma_s(g_-)$ (expressed as covariances in the 
Majorana representation) is given by~\cite{BGZ14}
\begin{equation}
F = \frac{\det[\id + \sqrt{
    e^{\tanh^{-1}{\Gamma_s(g_+)}} 
    e^{2\tanh^{-1}{\Gamma_s(g_-)}}
    e^{\tanh^{-1}{\Gamma_s(g_+)}}}
]^{1/2}}{
    \det[\id + e^{2\tanh^{-1}{\Gamma_s(g_+)}}]^{1/4}
    \det[\id + e^{2\tanh^{-1}{\Gamma_s(g_-)}}]^{1/4}
}.
\end{equation}

\begin{figure}[t]
\includegraphics[width=\columnwidth]{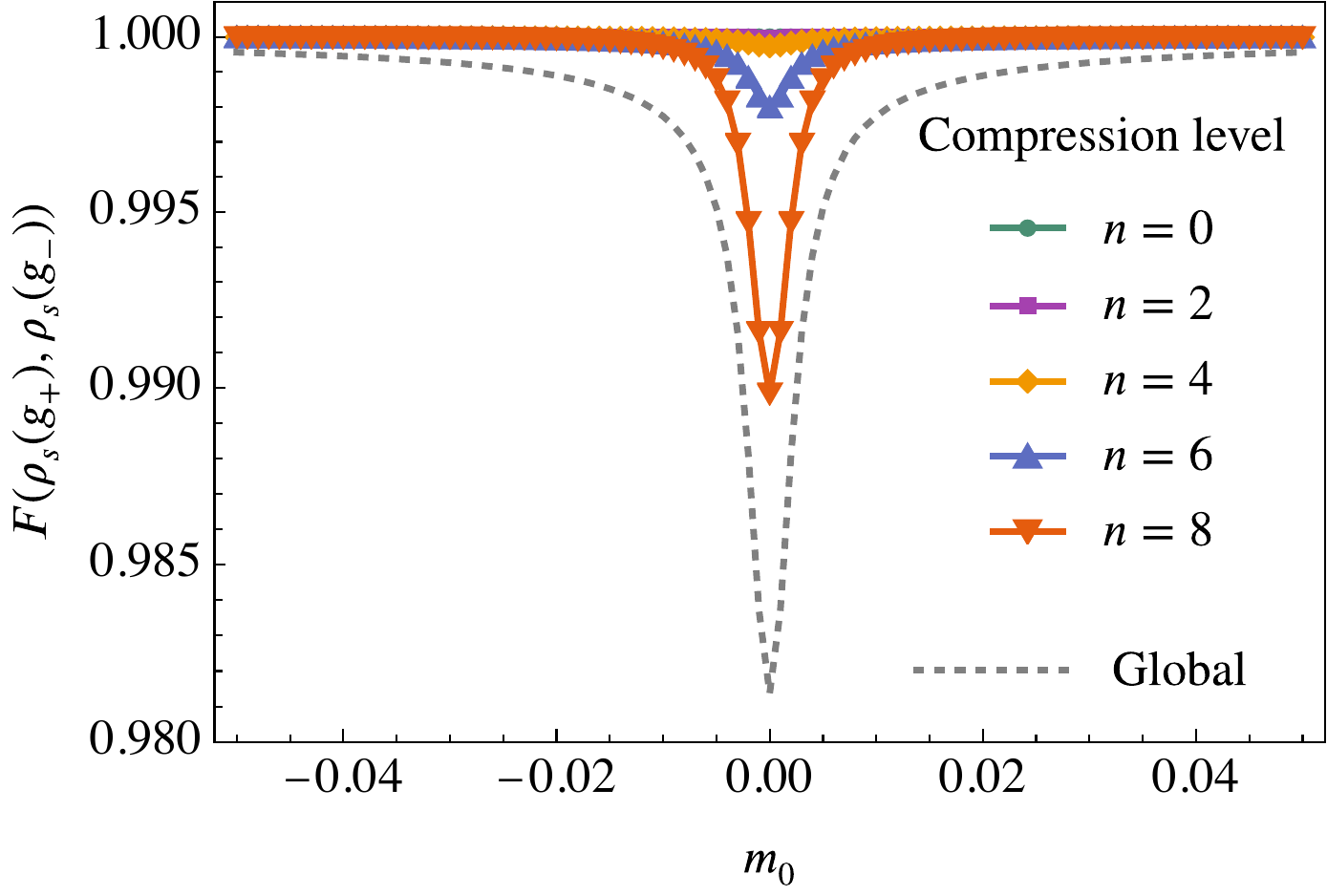}
\caption{\label{fig:results/fermionic_fidelity}
Fidelity of a 2-mode subsystem of scale modes at scale $r=0$ 
in a multiscale representation of the ground state of the 
fermionic QFT as a function of wavelet transform level $n$.
Here we keep $V=1024$ and $\delta=1/V$ constant.
Daubechies $\dbK=3$ wavelets were used.
The dashed grey line is the analytic expression for the fidelity
overlap of the global state
from \cref{eqn:analytic_fidelity_overlap}.
The phase transition at $m_0=0$ is increasingly evident for higher
compression levels.
}
\end{figure}

In \cref{fig:results/fermionic_fidelity} the signature
of a QPT is evident in the fidelity overlap calculated between 
two-mode subsystems of coarse modes. 
As expected, the minimum occurs at $m_0=0$ and is considerably 
more significant for subsystems of coarser modes (corresponding to 
higher wavelet transform levels) compared to finer modes.

We note also a dependence on the Daubechies wavelet index~$\dbK$, 
such that higher wavelet indices provide a better approximation
to the whole-state/analytic behavior with only a few coarse modes
(i.e. a lower fidelity minimum). 
However in the case of calculating the fidelity overlap the effect 
is weak, of order $10^{-5}$, and becomes weaker for higher $\dbK$.

%---------------------------------------------------------------------------
\subsection{Holographic entanglement of purification}
\label{sec:results/holographic_entanglement}

\begin{figure}
    \includegraphics[width=\columnwidth]{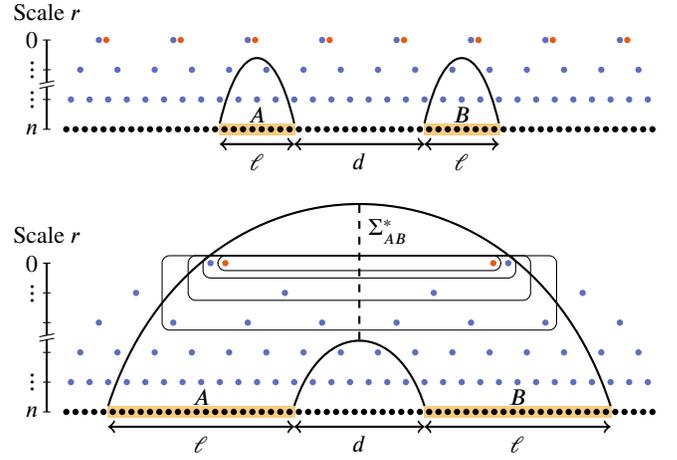}
    \caption{\label{fig:wedge_diagram}
    Illustration of the holographic entanglement features studied 
    in this paper for a ground state of a $(1+1)$D CFT with 
    periodic boundaries. 
    The physical degrees of freedom at the finest scale $r=n$ 
    are represented by scale modes (black). 
    Coarser wavelet modes at scale $r=n-1,\ldots,0$ 
    are shown (blue) as are the coarse scale modes at scale 
    $r=0$ (orange). 
    The representation of the ground state using all the wavelet 
    and coarse scale modes is referred to as the bulk description, 
    and is related by a unitary wavelet transformation to the 
    boundary description using scale modes at scale $n$. 
    Top: the size~$\ell$ of the regions $A,B$ is too small relative 
    to their separation $d$ so that $I(A\!:\!B)=0$. 
    There is no entanglement wedge and $E_W(\rho_{AB})=0$. 
    Bottom: here $d/\ell>\sqrt{2}-1$ and there is an entanglement wedge 
    with minimal cross section 
    $|\Sigma^*_{AB}|=E_W(\rho_{AB})=\frac{c}{6}\log(1+2d/\ell)$. 
    Shown in the smallest ellipse is a bulk subsystem involving only 
    coarse scale modes at scale $r=0$, a larger ellipse 
    involving scale and wavelet modes at scale $r=0$, and so 
    forth. 
    If the subsystems in these smaller sized ellipses accurately 
    capture the mutual information $I(A:B)$ with only a few scales, 
    then we speak of a compressed representation of $\rho_{AB}$.}
\end{figure}

Recently there has been progress in connecting entanglement in a 
boundary quantum field theory to geometric quantities in the bulk 
dual \cite{BTU18,UT18,Lee17}. 
A particularly compelling idea inspired by holographic duality 
is the conjectured equality \cite{UT18}
\begin{equation}
    E_p(\rho_{AB}) \stackrel{?}{=} E_W(\rho_{AB}),
\label{eqn:eopconjecture}
\end{equation}
where $E_p(\rho_{AB})$ is the entanglement of purification of a 
subsystem $\rho_{AB}$ of a boundary CFT, and $E_W(\rho_{AB})$ is 
the entanglement wedge cross section, a geometric quantity
defined in the bulk. 
The entanglement of purification is defined as
\begin{equation}
    E_p(\rho_{AB}) = {\rm min}_{
    \ket{\psi}_{A\bar{A}B\bar{B}};
    \Tr_{\bar{A}\bar{B}}[\ket{\psi}\bra{\psi}]=\rho_{AB}
    } 
    S(\rho_{A\bar{A}}),
\end{equation}
where $\bar{A},\bar{B}$ are auxiliary systems to $A, B$ and 
the minimum is taken over all purifications~$\ket{\psi}$ of the 
state~$\rho_{AB}$. 
The entanglement wedge cross section is
\begin{equation}
    E_W(\rho_{AB}) = \frac{\abs{\Sigma^{\ast}_{AB}}}{4G_N},   
\end{equation}
where $|\Sigma^{\ast}_{AB}|$ is the area of the minimal cross section 
of the entanglement wedge in the bulk dual that connects the boundary 
subsystem $A$ with $B$. 
We use units where $4G_N=1$. 
In the case of a $(1+1)$D boundary CFT, $\Sigma^{\ast}_{AB}$ is a 
one-dimensional surface. 

An appealing feature of the entanglement wedge cross section is that it is 
blind to cutoff-dependent features of the entanglement entropy due 
to cancellation of such terms.
The motivation for the conjectured equivalence in 
\cref{eqn:eopconjecture} is that---assuming that the global state 
is pure---the entanglement wedge satisfies several inequalities 
shared by the entanglement of purification (EoP) including
\begin{enumerate}
\item 
$I(A:B)/2 \leq E_W(\rho_{AB}) 
    \leq {\rm min} (S(\rho_\text{A}), S(\rho_\text{B}))$
\item
$E_W(\rho_{AB}) \leq E_W(\rho_{A(BC)}) 
    \leq E_W(\rho_{AB})+E_W(\rho_{BC})$
\item
$E_W(\rho_{(AA')(BB')}) \geq E_W(\rho_{AB}) + E_W(\rho_{A'B'})$  
\end{enumerate}
The first statement refers to the mutual information
\begin{equation}
I(A:B) = S(\rho_\text{A}) + S(\rho_\text{B}) - S(\rho_{AB}).  
\end{equation}
The left-hand side of the second statement follows from the 
extensiveness of the entanglement of mutual information and 
the right-hand side is a polygamy inequality. 
Both these inequalities are shared by EoP. 
The third inequality is a statement of strong superadditivity~\cite{UT18}. 
EoP in fact satisfies subadditivity: 
$E_P(\rho\otimes\sigma)\leq E_P(\rho)+E_P(\sigma)$ with equality 
only if the optimal purification of the joint state $\rho\otimes \sigma$ 
is the product of optimal purifications of $\rho$ and $\sigma$ 
separately, which is expected 
in holographic CFTs \cite{UT18}, in which case EoP becomes additive 
like $E_W$.

In order to better understand holographic entanglement of purification 
from a wavelet perspective, consider the case of a ground state 
$(1+1)$D CFT of overall length~$l$ and with periodic boundaries. 
The entanglement wedge cross section has an analytic formula, 
whereas the entanglement of purification does not, rather it 
involves a complex minimization.
To see the former, consider two regions $A$ and $B$, ordered left to 
right, separated by a distance $d$ and with equal lengths 
$\abs{A}=\abs{B}=\ell\ll L$, and with boundary points 
$\{\partial A_L,\partial A_R,\partial B_L,\partial B_R\}$. 
The entropy (up to an additive constant) of the joint region 
$A\cup B$ is given by the minimal length curve in AdS space that 
separates it from its complement.
This will either be the sum of the geodesic connecting boundary 
points $\{\partial A_L,\partial A_R\}$ and that connecting
$\{\partial B_L,\partial B_R\}$ (see \cref{fig:wedge_diagram}, top), 
i.e. $\frac{2c}{3}\log(\ell)$, or the sum of the geodesic 
which connects $\{\partial A_L,\partial B_R\}$ and that 
connecting $\{\partial A_R,\partial B_L\}$ (see
\cref{fig:wedge_diagram}, bottom).
i.e, $\frac{c}{3}\log(d(d+2\ell))$.
These two lengths become degenerate at $d/\ell = \sqrt{2}-1$.
The mutual information is then
\begin{equation}
I(A:B) = \begin{cases}
    \hfil 0 & d/\ell \geq \sqrt{2}-1 \\
    -\frac{c}{3}\log((d/\ell)^2+2d/\ell) & d/\ell < \sqrt{2}-1
\end{cases}. 
\end{equation}
For regions $A,B$ satisfying $I(A:B)>0$, the entanglement 
wedge~$W$ is the region in the bulk bounded by $A\cup B$ and the 
minimal length geodesics separating $A\cup B$ from its complement 
(\cref{fig:wedge_diagram}, bottom).  
The entanglement wedge cross section $\abs{\Sigma^*_{AB}}$ is the 
length of the shortest curve $\Sigma_{AB}$ in AdS space that divides 
the wedge into two pieces: one containing $A$, the other $B$.
Because the AdS distance from the midpoint of a geodesic to the 
boundary point bisecting the boundary curve is one half the length 
of the geodesic, the entanglement wedge cross section is therefore
(for the example above)
\begin{equation}
    E_W(\rho_{AB}) = \frac{c}{6}\log(1+2\ell/d).
\label{eqn:ewedge}
\end{equation}
In terms of the bulk modes, we seek to compute the 
entanglement of purification $E_p(\rho_{AB})$ via a compressed 
representation of $\rho_{AB}$ only involving a few coarse scale 
and wavelet modes. 
As illustrated in \cref{fig:wedge_diagram} (bottom), if small 
bulk subsystems accurately capture the mutual information $I(A:B)$ 
then it will suffice to restrict to this small compressed subsystem 
to calculate the entanglement of purification. 
Ideally, if the state $\rho_{AB}$ is represented only in terms of 
two coarse scale modes and two coarse wavelet modes both at 
scale~$l_{\rm min}$ then it may be possible to analytically 
compute~$E_P(\rho_{AB})$.

\begin{figure}
\includegraphics[width=\columnwidth]{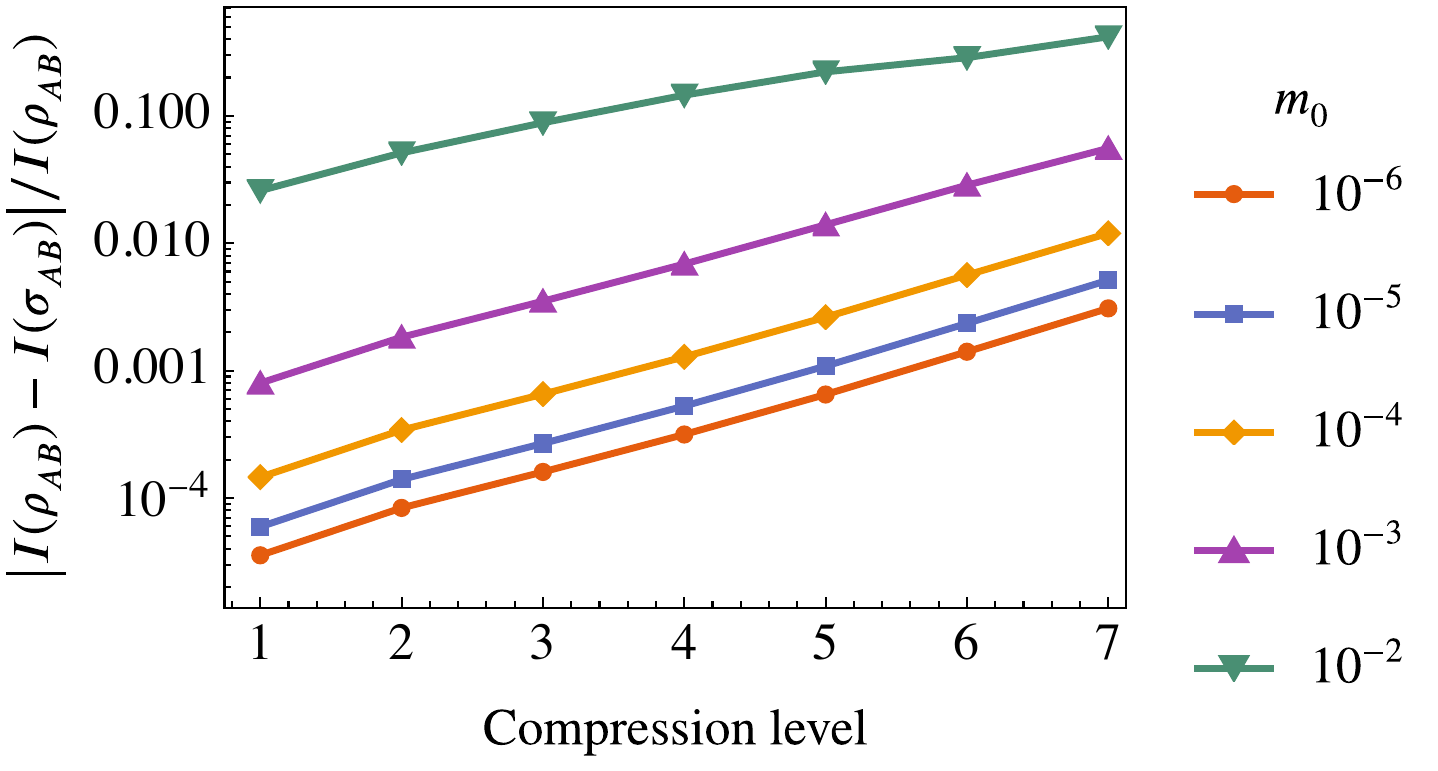}
\caption{\label{fig:mutual_info_rel_error}
Relative error of approximating the mutual information $I(A:B)$ by the 
wavelet compressed representation $\sigma_{AB}$ of the 
reduced state $\rho_{AB}$, where $A$ and $B$ are subsystems of the 
ground state of a $(1+1)$-dimensional scalar bosonic field theory 
with a total system size of $V=4096$.
The subsystems are chosen to have a separation of inner 
boundaries by a length $d=512$ and are of equal size 
$\ell=\abs{A}=\abs{B}=256$.
For each additional compression the number of degrees of 
freedom is halved.
}
\end{figure}

\begin{figure}
\includegraphics[width=\columnwidth]{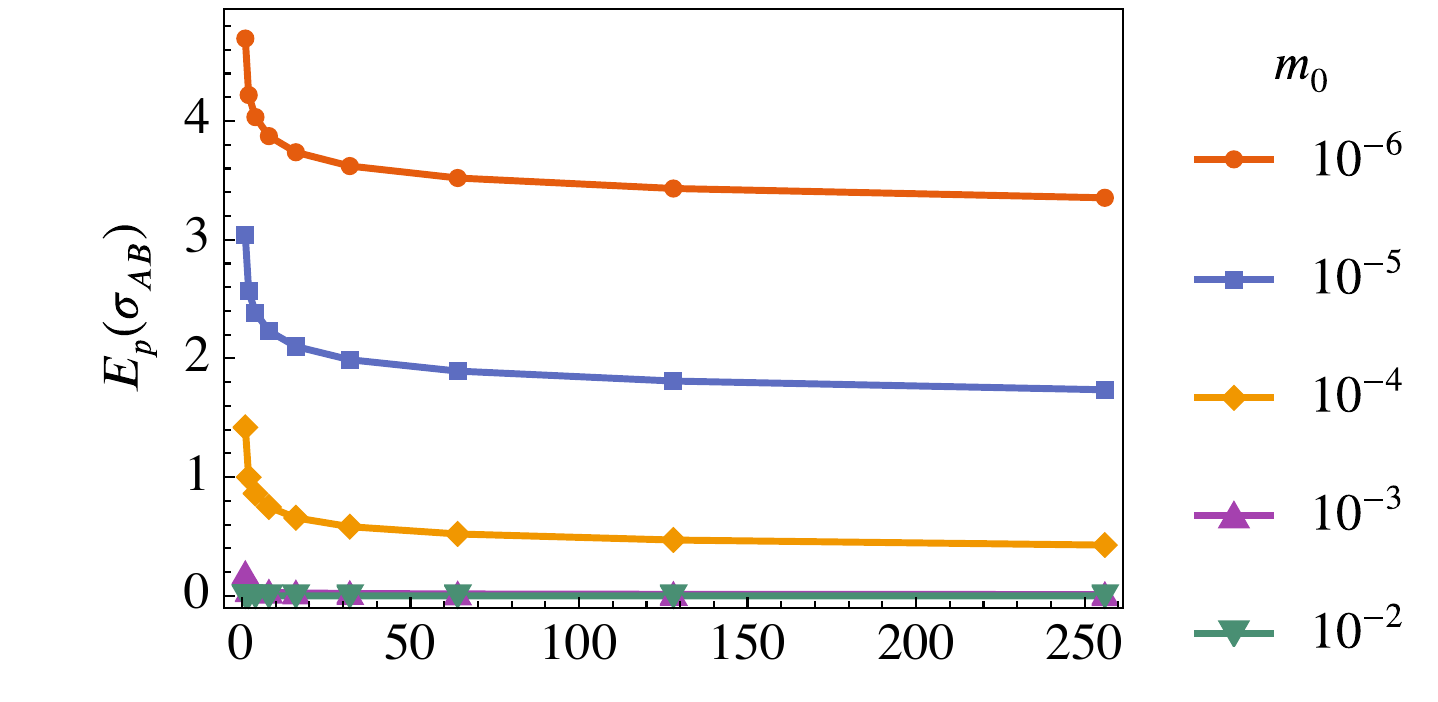}
\includegraphics[width=\columnwidth]{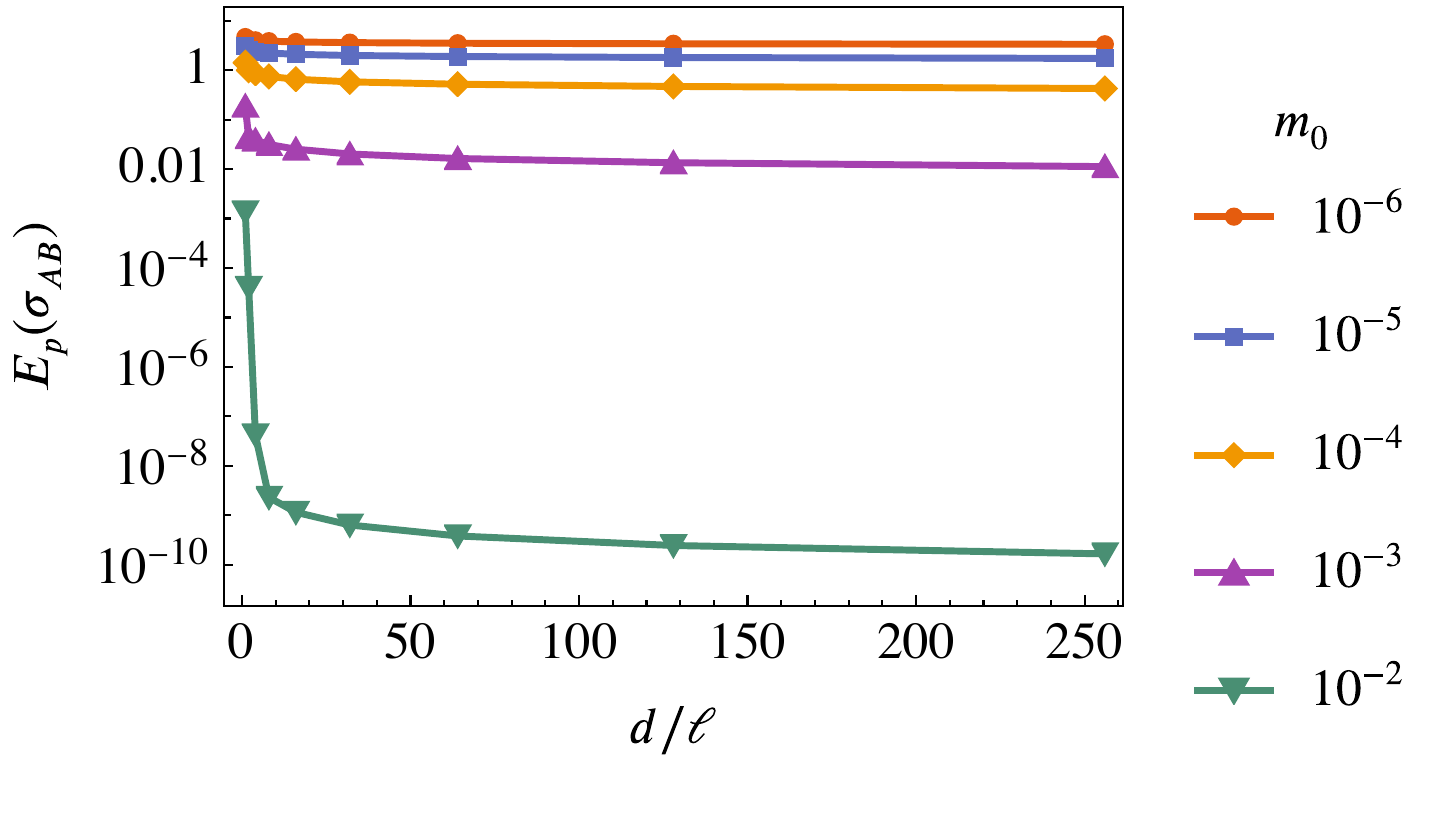}
\caption{\label{fig:entanglement_of_purification}
Top: entanglement of purification $E_p(\sigma_{AB})$ as a function
of the ratio $d/\ell$ as captured by wavelet compression of the 
ground state as in \cref{fig:mutual_info_rel_error}.
Here $\sigma_{AB}$ is the reduced state represented by two
coarse grained scale modes obtained by the wavelet transformation 
on the covariance matrix representation of $\rho_{AB}$. 
The separation between subsystems $d=512$ is fixed, and
$\ell=\abs{A}=\abs{B}$ is varied from $512$ to $2$.
The compressed state allows for a calculation of entanglement 
of purification in terms of a one parameter minimization.
Bottom: same as top but on a log scale. Compare with 
\textcite{BTU18} (Fig. 6 there).
}
\end{figure}

\begin{figure}
\includegraphics[width=\columnwidth]{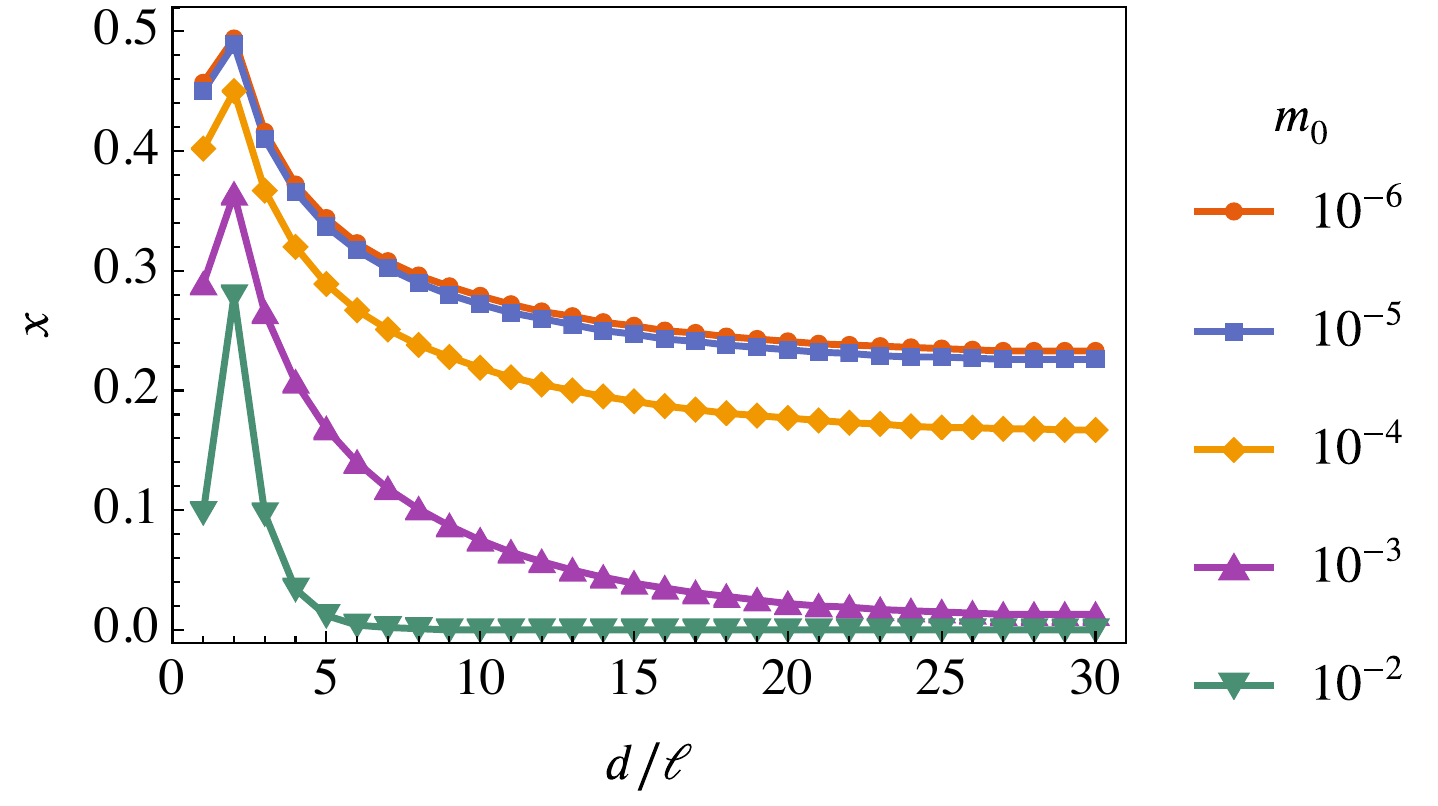}
\caption{\label{fig:eop_phase_transition}
Minimizing value $x$ for the entanglement of purification of a 
compressed ground state as in \cref{fig:entanglement_of_purification} 
plotted as a function of $d/\ell$, focusing on the 
region $d/\ell \leq 30$.
The peak in the value at $d/\ell=2$ is indicative of a phase 
transition near that ratio.
}
\end{figure}

Following the scheme in \cref{fig:wedge_diagram} (bottom) we study 
some examples 
of subsystem sizes~$\ell$ and separations $d$ that have nonzero 
mutual information calculated using the boundary scale modes at 
scale $n$, under the assumption $\ell,d\ll L$.
By computing a wavelet transform on the covariance matrix we find 
cases where $I(A:B)$ is accurately represented by a few coarse 
scale and wavelet modes, as demonstrated in 
\cref{fig:mutual_info_rel_error}. 

Assume there exists a state for which $I(A:B)$ is well approximated 
up to some small additive error by a compression to two coarse scale 
modes (one for $A$ and one for $B$), and let the the reduced two-mode 
state be denoted~$\sigma_{AB}$.
Calculating $E_P(\sigma_{AB})$ following the method of
\citeauthor{BTU18} (Sec. 4.1 in \textcite{BTU18}) requires minimizing 
the entropy~$S(\sigma_{A\bar{A}})$ over a single real parameter.
The corresponding minimization for a state compressed to two coarse 
scale modes and two coarse wavelet modes (see Sec. 4.3 in \cite{BTU18}) 
requires minimizing the entropy~$S(\sigma_{A\bar{A}})$ over four real 
parameters.
Both these methods assume the minimizing pure state is also Gaussian.
This greatly simplifies the analysis, as we can continue to represent 
states with covariance matrices.
The results, plotted in \cref{fig:entanglement_of_purification}, 
appear to validate this assumption.
The behavior of $E_P(\sigma_{AB})$ is comparable to \cref{eqn:ewedge},
the analytic formula for the entanglement wedge cross section
$E_W(\rho_{AB})$.

A phase transition also becomes apparent in the value of the 
single minimization parameter $x$ during the process of minimizing
for the entropy $S(\rho_{A\bar{A}})$.
Following the approach in \textcite{BTU18}, consider a pure state on a 
total system $A\bar{A}B\bar{B}$ with covariance matrices
\begin{equation}
\Gamma^{\Pi\Pi}_{AB\bar{A}\bar{B}} = \frac12
\begin{pmatrix}
    J & K \\
    K^T& L
\end{pmatrix}, \quad
\Gamma^{\Phi\Phi}_{AB\bar{A}\bar{B}} = \frac12
\begin{pmatrix}
    D & E \\
    E^T & F
\end{pmatrix}
\end{equation}
such that
\begin{equation}
\begin{pmatrix}
    J & K \\
    K^T& L
\end{pmatrix}^{-1} = \begin{pmatrix}
    D & E \\
    E^T & F
\end{pmatrix}.
\label{inverses}
\end{equation}
The matrices are written in the 
basis~$(\Phi_{AB},\Phi_{\bar{A}\bar{B}})$, where the known reduced 
state covariance matrices 
are~$\Gamma^{\Pi\Pi}_{AB}=\frac{1}{2}J$ 
and~$\Gamma^{\Phi\Phi}_{AB}=\frac{1}{2}D$.
Now it follows from \cref{inverses} that
\begin{equation}
\begin{split}
JD+KE^T = K^TE+LF &= \id,
\\
JE+KF = K^TD+LE^T &= 0,
\end{split}
\end{equation}
which implies $L=-K^TD(\id-JD)^{-1}K$.
Hence given the covariance matrix~$\Gamma_{AB}$ the purified 
state covariance matrix is completely specified by $K$. 
The dimensions of~$K$ will depend on the size of the auxiliary 
spaces~$\bar{A}\bar{B}$. 
\textcite{BTU18} show that for the case where $\abs{A}=\abs{B}=1$, an 
accurate value of the entanglement of purification can be obtained 
by choosing~$\abs{\bar{A}}=\abs{\bar{B}}=1$, meaning that the value 
obtained is negligibly changed by choosing larger auxiliary systems.
Furthermore, by invoking exchange symmetry of $A$ and $B$, $K$ can be 
chosen to have the canonical form
\begin{equation}
K = \begin{pmatrix}
    1 & x \\
    x & 1
\end{pmatrix}
\end{equation}
where $x\in (-1,1)$.
The value of $x$ should be selected in order to minimize the entropy 
of the reduced state of $A\bar{A}$ described from the reduced 
covariance matrix $\Gamma_{A\bar{A}}$ which is found by tracing out 
rows and columns of $\Gamma_{AB\bar{A}\bar{B}}$.

A plot of the value of the parameter~$x$ that minimizes the entropy 
of the reduced state $A\bar{A}$ is shown in \cref{fig:eop_phase_transition} 
for several decades of mass. There is a notable peak in the value 
of~$x$ at the subregion size to separation distance ratio~$d/\ell=2$, 
indicative of a phase transition near that value.

%---------------------------------------------------------------------------
%   DISCUSSION
%---------------------------------------------------------------------------
\section{Discussion}
\label{sec:discussion}

We show in result \ref{sec:results/correlators} that the same-scale 
correlators in the multiscale representation of the ground states 
for the bosonic and fermionic theories demonstrate the expected 
power-law decay in the massless case, with an exponent that depends 
on the Daubechies index~$\dbK$, and exponential decay in the massive 
cases.
Mass renormalization is naturally emergent as a function 
of scale. 
We also consider in result \ref{sec:results/subsystem_entropy} the
entanglement features of the ground states of the 
two QFTs in a scale field representation and verified that the 
Calabrese-Cardy relations are obeyed.
For the massive bosonic theory in 1D and 2D, we observe that the 
entanglement entropy of a subsystem is constant as a function of 
the subsystem's length in different scales. 
The constant value increases as we increase the scale 
parameter~$k$ as is to be expected from the entanglement area law. 
For the massless bosonic and fermionic CFTs, we obtain the correct 
central charges and also the cutoff dependence of the entropy as a 
function of scale.

Results \ref{sec:results/phases} and 
\ref{sec:results/holographic_entanglement} demonstrate two potential
applications of using a wavelet-based multiscale representation as 
a form of compression, where some function on a system with $2^n X$ 
modes can be approximated by applying that function to a reduced state
of $X$ coarse scale modes obtained from an $n$-level wavelet transform.

In the case of result \ref{sec:results/phases}, we show that in a 
fermionic QFT this wavelet compression technique can be used to identify 
a phase transition, evident in a decrease in fidelity overlap 
between ground states adjacent in some parameter space (here, mass).
Given that the fast wavelet transform (FWT) has an efficient classical 
implementation that scales with $\mathcal{O}(N \log(N))$,
where~$N$ is the dimension of the vector space, this technique
holds promise for approximating the value of a many-body observable
that might otherwise be prohibitively difficult to observe directly 
due to experimental, computational, or other constraints.

Finally, result \ref{sec:results/holographic_entanglement} demonstrates 
that, for a bosonic QFT, wavelet compression qualitatively captures 
the physics of the mutual information between isolated subsystems, 
including identification of a phase transition.

The use of higher-order wavelet basis functions results in more desirable
mathematical properties, such as increased accuracy of approximations 
using a small number of modes, and well-defined higher-order 
derivatives, at the cost of moderately increased computational complexity.
As noted in \textcite{BK97} (page 179), the error incurred by the 
wavelet discretization method on a second-order differential 
equation described by a Laplacian is $\order{(\Delta x)^{2\dbK}}$ 
where $\Delta x = 2^{-k}$ and $k$ is the number of scales in the
multiresolution analysis.
We can extrapolate this point to the bosonic field theory for an
$\order{2^{-2k\dbK}}$ error scaling, and to the fermionic field theory 
(arises from a first-order equation) to find an error scaling of 
$\order{2^{-k\dbK}}$.
This suggests that we have a strategy for reducing error in the 
discretization that is not simply increasing the number of 
scales~$k$, i.e.~reducing the size of the length cutoff. 
We may also reduce error by increasing the value of $\dbK$. 
Note however that increasing $\dbK$ results in a corresponding 
computational cost since the number of nonzero bands in the 
associated wavelet transform scales as $2\dbK$.

In this paper we have mostly used a uniform wavelet basis with
either periodic or antiperiodic boundary conditions.
This conforms to the usual application of the discrete wavelet 
transform with periodic/antiperiodic signal extension modes in 
numerical signal processing.
In the case of \cref{sec:results/subsystem_entropy}, the application
of open boundaries corresponds to the absence of a signal extension, 
which is sufficient for the calculation of bulk entanglement since
this property is sensitive primarily to the underlying topology of 
the space.
However, when studying open or nonperiodic systems with essential 
physics at the boundaries, for example, symmetry-protected 
topological phases~\cite{CGLW13}, then a careful consideration of 
the wavelet representation of that theory at the boundaries is 
necessary.
For an overview of boundary wavelet construction, 
see~\textcite{Mal09}(pages 322-328).

%---------------------------------------------------------------------------
%   CONCLUSION
%---------------------------------------------------------------------------
\section{Conclusion}
\label{sec:conclusion}

We have demonstrated the utility of wavelet analysis
when describing quantities such as entanglement in fermionic and
bosonic QFTs.
Specifically, the scale dependence of various quantities such as 
subsystem entropy and correlations emerge simply by fixing 
a wavelet basis, unlike, for example, tensor network 
representations, where generally the elements of the tensors 
must be obtained by numerical optimization. 
We have additionally shown that wavelets provide a way to compress 
quantum states in a way that enables the calculation of quantum
informational quantities on a very few number of modes. 
Such a result could be useful in experimental probes of quantum 
simulations of QFTs where measuring observables over an extensive 
number of modes is costly or error prone.

Wavelet analyses have potential in more general QFT 
simulation algorithms, and are already showing promise in algorithms
for ground state generation with spatial inhomogeneities~\cite{BSB+22}.
By showing that cutoff-dependent results like those of \textcite{CC04} 
appear directly as a function of an input scale parameter in 
wavelet-based representations of QFT, we bolster the case for 
wavelet-based representations as a key tool of analysis for the 
physics of quantum fields.

%---------------------------------------------------------------------------
%   ACKNOWLEDGMENTS
%---------------------------------------------------------------------------
\begin{acknowledgments}
Y. R. S. thanks Eric Howard for many helpful discussions on lattice
regularisation and entanglement area laws.
G. K. B. acknowledges helpful discussions with Dean Southwood.
D. J. G., Y. R. S., and G. K. B. acknowledge the Wallamattagal people of the 
Dharug nation, whose cultures and customs have nurtured, and 
continue to nurture, the land on which some of this work was 
undertaken: Macquarie University.
M. B. and B. C. S. acknowledge the traditional owners of the land on which 
some of this work was undertaken at the University of Calgary: the 
Treaty 7 First Nations.
D. J. G. and G. K. B. acknowledge support from the Australian Research 
Council (ARC) through Grant No. DP200102152 and from the ARC Centre of 
Excellence for Engineered Quantum Systems (EQUS, CE170100009).
D. J. G. was supported by the Sydney Quantum Academy, Sydney, Australia.
M. B. and B. C. S. acknowledge support from the Government of Alberta and 
by the Natural Sciences and Engineering Research Council of Canada (NSERC).

\end{acknowledgments}

%---------------------------------------------------------------------------
%   BIBLIOGRAPHY
%---------------------------------------------------------------------------
\bibliography{main.bib}

%---------------------------------------------------------------------------
%   APPENDICES
%---------------------------------------------------------------------------
\appendix

%---------------------------------------------------------------------------
%   FERMIONIC CORRELATIONS
%---------------------------------------------------------------------------
\section{Expressions for fermionic covariance matrix}
\label{sec:fermionic_correlations_derivation}

%---------------------------------------------------------------------------
\subsection{General solution in wavelet scale basis}

The Hamiltonian for the Ising model field theory in the continuous case is
\begin{equation}
    \hat{H} = \frac{1}{2} \int \! \dd{x} \; \left[\,
        \hat{\bm{c}}^\dag(x) \, \ii \, Y \, \partial_x \, \hat{\bm{c}}(x)
        + m_0 \, \hat{\bm{c}}^\dag(x) \, Z \, \hat{\bm{c}}(x) \,
    \right],
\end{equation}
where  
$\ii \, Y = \begin{pmatrix} 0 & 1 \\ -1 & 0 \\ \end{pmatrix}$,  
$Z = \begin{pmatrix} 1 & 0 \\ 0 & -1 \\ \end{pmatrix}$
and the fermionic field operators are 
$\hat{\bm{c}}(x) = \begin{pmatrix} 
    \hat{c} (x) \\ \hat{c}^\dag (x) \\ \end{pmatrix}$, 
$\hat{\bm{c}}^\dag (x) = \begin{pmatrix} 
    \hat{c}^\dag (x) & \hat{c}(x) \\ \end{pmatrix}$, with the
anticommutation relation 
$\left\{ \hat{c} (x), \hat{c}^\dag (x') \right\} = \delta (x - x')$. 
Discretizing the continuous Hamiltonian into $2^n X = V$ scale modes 
$\hat{\bm{r}}_\ell = \begin{pmatrix} 
    \hat{r}_\ell \\ \hat{r}^\dag_\ell \\ \end{pmatrix}$ 
where  
$\hat{r}_\ell = \hat{r}_\ell^{(n;s)} = 
    \int \! \dd{x} \; s^{(n)}_\ell (x) \, \hat{c} (x)$ and 
$\left\{ \hat{r}_\ell, \hat{r}_{\ell'}^\dag \right\} = 
    \delta_{\ell,\ell'}$ gives the discrete Hamiltonian in the 
scale-n basis
\begin{equation}
    \hat{H}^{(n)} = \frac{1}{2} \sum_{\ell=0}^{V-1} \left[
        \sum_{j = -2\dbK+2}^{2\dbK-2}
        \hat{\bm{r}}_\ell^\dag \, \ii \, Y \, \Delta_j^{(1)} \,
        \hat{\bm{r}}_{\ell+j} +  m_0 \, \hat{\bm{r}}_\ell^\dag \, Z \,
        \hat{\bm{r}}_\ell 
    \right],
\end{equation}
where $\dbK$ is the Daubechies wavelet index and 
$\Delta_j^{(1)} \equiv \Delta_{\ell,\ell+j}^{(1)}$ is the first 
derivative operator in the base scale (scale-0 as
per \cref{eqn:wavelets/scale_notation}) and is nonzero only when 
$-2\dbK+2 \leq j \leq 2\dbK-2$.

Note that $\hat{H}$ commutes with the total fermionic parity operator 
$\hat{J} = \prod_\ell \left( 1 - 2\hat{r}_\ell^\dag \hat{r}_\ell \right)$, 
and in order to have translational invariance on the even parity 
sector, application of antiperiodic boundary conditions requires  
$\hat{r}_{V+\ell} \equiv -\hat{r}_\ell$. 
In practice this means that the upper-right and lower-left corner 
terms in the matrix $\Delta^{(1)}$ will be the negative of those 
along the corresponding main diagonals.

This Hamiltonian can be expressed in terms of uncoupled 
modes~$\hat{\eta}_k$ in the usual diagonal form as follows:
\begin{equation}
\hat{H} = \sum_{k \in S} \omega_k \left( 
    \hat{\eta}_k^\dag \hat{\eta}_k - \frac{1}{2} 
\right)
\end{equation}
where 
$S = \left\{ \frac{1}{2}, \frac{3}{2}, \cdots, V - \frac{1}{2} \right\}$
due to the antiperiodic boundary conditions.
The ground state $\ket{G}$ is defined by 
$\hat{\eta}_k \ket{G} = 0$, from which follow the uncoupled correlations  
$\ev{\hat{\eta}_k \hat{\eta}_{k'}} 
    = \ev{\hat{\eta}_k^\dag \hat{\eta}_{k'}^\dag} 
    = \ev{\hat{\eta}_k^\dag \hat{\eta}_{k'}} = 0$, 
and $\ev{\hat{\eta}_k \hat{\eta}_{k'}^\dag} = \delta_{k,k'}$.

The uncoupled modes $\hat{\eta}_k$ are related to the original 
fermionic modes $\hat{r}_\ell$ via a pair of transforms. 
The original modes are related to the momenta modes by the usual 
Fourier transform 
$\hat{p}_k = \frac{1}{\sqrt{V}} 
    \sum_{\ell=0}^{V-1} \hat{r}_\ell \ee^{\ii 2 \pi k \ell / V}$.
The momenta modes are related to the uncoupled modes by way of the 
Bogoliubov transform
$\hat{\eta}_k = u_k \hat{p}_k + \ii v_k \hat{p}_{-k}^\dag$, where 
$u_k = -(m_0 + \omega_k)/\sqrt{(m_0 + \omega_k)^2 + q_k^2}$, 
$v_k = q_k/\sqrt{(m_0 + \omega_k)^2 + q_k^2}$,   
$\omega_k=\sqrt{m_0^2 + q_k^2}$  ($\omega_k$ are the eigenvalues of 
the Hamiltonian) and $q_k = 2 \sum_{j=1}^{2\dbK-2} \Delta^{(1)}_j 
    \sin{\frac{2 \pi j k}{V}}$.
The combined transform can be expressed in the form:
\begin{align}
\hat{\eta}_k &= 
    \frac{u_k}{\sqrt{V}} \sum_{\ell=0}^{V-1} 
        \hat{r}_\ell \ee^{\ii 2 \pi k \ell / V} + 
    \frac{\ii v_k}{\sqrt{V}} \sum_{\ell=0}^{V-1} 
        \hat{r}_\ell^\dag \ee^{\ii 2 \pi k \ell / V}.
\end{align}

Introduce the Majorana scale modes $\hat{\bm{b}}_\ell = \begin{pmatrix} 
\hat{b}_{\ell,0} \\ \hat{b}_{\ell,1} \\ \end{pmatrix}$
with $\hat{b}_{\ell,0} = \hat{b}_{\ell,0}^{(n;s)} = 
\hat{r}_\ell + \hat{r}_\ell^\dag$, 
$\hat{b}_{\ell,1} = \hat{b}_{\ell,1}^{(n;s)} 
    = -\ii \left( \hat{r}_\ell - \hat{r}_\ell^\dag \right)$, such that 
$\left\{ \hat{b}_{\ell,\sigma}, \hat{b}_{\ell',\sigma'} \right\} 
    = 2 \delta_{\ell, \ell'} \delta_{\sigma, \sigma'}$.
The Hamiltonian transforms to
\begin{equation}
\hat{H}^{(n)} = -\frac{1}{4} \sum_{\ell=0}^{V-1}
    \left[ \sum_{j=-2\dbK+2}^{2\dbK-2} \, 
    \hat{\bm{b}}_\ell^T \, \ii \, X \, 
        \Delta_j^{(1)} \, \hat{\bm{b}}_{\ell+j} 
    + m_0 \, \hat{\bm{b}}_\ell^T \, Y \,\hat{\bm{b}}_\ell \right].
\end{equation}

That $u_k = u_{-k}$ and $v_k = -v_{-k}$ follows from the properties 
of the Bogoliubov transform.
Furthermore introduce $\theta_k$ such that 
$u_k = \cos{\theta_k}$, $v_k = \sin{\theta_k}$, and so
\begin{align}
\begin{split}
\hat{\eta}_k + \hat{\eta}_{-k}^\dag 
&= \frac{\ee^{\ii \theta_k}}{\sqrt{V}} \sum_{\ell=0}^{V-1} 
    \hat{b}_{\ell,0} \ee^{\ii 2 \pi k \ell / V}
\\ 
\hat{\eta}_k - \hat{\eta}_{-k}^\dag 
&= \frac{\ii \ee^{-\ii \theta_k}}{\sqrt{V}} \sum_{\ell=0}^{V-1} 
    \hat{b}_{\ell,1} \ee^{\ii 2 \pi k \ell / V}
\end{split}
\\
\begin{split}
\hat{b}_{\ell,0} 
&= \frac{1}{\sqrt{V}} \sum_{k \in S} 
    \ee^{-\ii \theta_k} (\hat{\eta}_k + \hat{\eta}_{-k}^\dag) \, 
    \ee^{-\ii 2 \pi k \ell / V}
\\ 
\hat{b}_{\ell,1} 
&= \frac{1}{\sqrt{V}} \sum_{k \in S} -\ii 
    \ee^{-\ii\theta_k} (\hat{\eta}_k - \hat{\eta}_{-k}^\dag) \, 
    \ee^{-\ii 2 \pi k \ell / V}
\end{split}
\end{align}

Noting that $\theta_{-k} = \arctan{(v_{-k}/u_{-k})} 
= \arctan{(-v_k/u_k)} = -\theta_{k}$, the correlations can now be 
computed directly:
\begin{align}
\begin{split}
\ev{\hat{b}_{\ell,0} \hat{b}_{\ell',0}} 
&= \frac{1}{V} \sum_{k,k' \in S} 
    \ee^{-\ii (\theta_k + \theta_{k'})} 
    \ee^{-\ii 2 \pi (k \ell + k' \ell')/V}
\\
& \qquad \times \underbrace{\ev{(\hat{\eta}_k + \hat{\eta}^\dag_{-k})
    (\hat{\eta}_{k'} + \hat{\eta}^\dag_{-k'})}}_{= \delta_{k,-k'}}
\\
&= \delta_{\ell,\ell'}
\end{split}
\\
\begin{split}
\ev{\hat{b}_{\ell,0} \hat{b}_{\ell',1}}
&= \frac{\ii}{V} \sum_{k,k' \in S} 
    \ee^{-\ii (\theta_k - \theta_{k'})} 
    \ee^{-\ii 2 \pi (k\ell + k'\ell')/V}
\\
& \qquad \times \underbrace{\ev{(\hat{\eta}_k + \hat{\eta}^\dag_{-k})
    (\hat{\eta}_{k'} - \hat{\eta}^\dag_{-k'})}}_{= -\delta_{k,-k'}}
\\
&= \frac{\ii}{V} \sum_{k \in S}
\ee^{-2\ii \theta_k} \ee^{-\ii 2 \pi (\ell - \ell') k / V}
\end{split}
\end{align}

Similarly 
$\ev{\hat{b}_{\ell,1} \hat{b}_{\ell',1}} 
    = \delta_{\ell,\ell'}$ 
and $\ev{\hat{b}_{\ell,1} \hat{b}_{\ell',0}} 
    = -\ev{\hat{b}_{\ell,0} \hat{b}_{\ell',1}}$. 
The covariance matrix $\Gamma$ defined as 
$\ev{\hat{b}_{\ell,\sigma} \hat{b}_{\ell',\sigma'}} 
    = \delta_{\ell,\ell'}\delta_{\sigma,\sigma'} 
    + \ii \Gamma_{\sigma,\ell;\sigma'\ell'}$ in the basis 
$\ket{\sigma} \ket{\ell}, \sigma \in \{0, 1\}$ is therefore
\begin{align}
\Gamma_{\sigma,\ell;\sigma',\ell'}
&= \begin{bmatrix}
    0 & \Gamma^\text{01} \\
    -(\Gamma^\text{01})^T & 0 \\
\end{bmatrix}
\label{eqn:fermionic_correlations/correlation_matrix}
\\
\Gamma_{\ell,\ell'}^\text{01}
&= \frac{1}{V} \sum_{k \in S} 
\ee^{-2 \ii \theta_k} \ee^{-\ii 2 \pi (\ell - \ell') k / V}
\label{eqn:fermionic_correlations/correlation_submatrix}
\end{align}
with $\theta_k = \arctan{\frac{-q_k}{m_0 + \sqrt{m_0^2 + q_k^2}}}$ and 
$q_k = 2 \sum_{j=1}^{2\dbK-2} \Delta^{(1)}_j \sin{\frac{2 \pi j k}{V}}$.

%---------------------------------------------------------------------------
\subsection{Zero-mass limit}

Consider the zero-mass limit $m_0 \to 0$:
\begin{align}
\lim_{m_0 \to 0} \theta_k
&= \lim_{m_0 \to 0} \arctan{\frac{-q_k}{m_0 
+ \sqrt{m_0^2 + q_k^2}}} \nonumber
\\
&= -\frac{\pi}{4} \sgn{q_k} \nonumber
\\
&= \begin{cases}
    \frac{\pi}{4} & k \in S_L \\
    -\frac{\pi}{4} & k \in S_U \\
\end{cases}
\end{align}
where $S_L$ is the lower half of momenta 
modes~$S_L: \left\{ \frac{1}{2}, \frac{3}{2}, \dots, 
    \frac{V-1}{2} \right\}$ and $S_U$ is the upper 
half~$S_U: \left\{ \frac{V+1}{2}, \frac{V+3}{2}, \dots, 
    \frac{2V-1}{2} \right\}$.
The covariance matrix is then
\begin{equation}
\begin{split}
\lim_{m_0 \to 0} \Gamma^\text{01}_{\ell,\ell'} 
= -\frac{\ii}{V} \Bigl[ &\sum_{k \in S_L}
    \ee^{-\ii 2 \pi (\ell - \ell') k / V} 
\\
- &\sum_{k \in S_U}
    \ee^{-\ii 2 \pi (\ell - \ell') k / V} \Bigr].
\end{split}
\end{equation}

Let $\zeta = \ee^{- \ii \pi (\ell - \ell') / V}$, then
\begin{align}
\begin{split}
\lim_{m_0 \to 0} \Gamma^\text{01}_{\ell,\ell'} 
&= -\frac{\ii}{V} \bigl[ 
    \left( \zeta + \zeta^3 + \cdots + \zeta^{V-1} \right)
\\
& \quad - \left( \zeta^{V+1} + \zeta^{V+3}
        + \cdots + \zeta^{2V-1} \right) \bigr]
\end{split}
\\
\begin{split}
\lim_{m_0 \to 0} \bigl( \zeta^2 \Gamma^\text{01}_{\ell,\ell'} 
    - &\Gamma^\text{01}_{\ell,\ell'} \bigr)
\\
&= -\frac{\ii}{V} \bigl[  -\zeta + \zeta^{V+1} + \zeta^{V+1} 
    - \zeta^{2V+1} \bigr] 
\end{split}
\\
\lim_{m_0 \to 0} \Gamma^\text{01}_{\ell,\ell'} 
&= -\frac{\ii}{V} \frac{1}{\zeta - \zeta^{-1}}
    \left( 2\ee^{-\ii \pi (\ell - \ell')} - 2 \right)
\\
&= \begin{cases}
    \frac{-2}{V \sin{ \left( \pi (\ell - \ell') / V \right) }} & 
    \ell-\ell' \text{ odd} \\
    0 & \ell-\ell' \text{ even} \\
    \end{cases}
\end{align}

%---------------------------------------------------------------------------
\subsection{Finite mass}

For finite mass, define $s_k = q_k/m_0$.
Then for $m_0 \gg \abs{q_k}$,
$\theta_k 
    = \arctan{\left( -s_k / \left( 1 + \sqrt{1 + s_k^2} \right) \right)} 
    \approx -s_k/2 + O(s_k^3)$, and so
\begin{equation}
\Gamma^\text{01}_{\ell,\ell'} 
= \frac{1}{V} \sum_{k \in S}
    \ee^{\ii s_k} \ee^{-\ii 2 \pi (\ell - \ell') k / V}.
\end{equation}
In the special case of Haar wavelets ($\dbK = 1$),
$\Delta^{(1)}_1 = -\frac{1}{2}$, $\Delta^{(1)}_{\ell>1} = 0$, and so 
$s_k = -\frac{1}{m_0} \sin{ \left( 2 \pi k / V \right)}$:
\begin{align}
\Gamma^\text{01}_{\ell,\ell'}
&= \frac{1}{V} \sum_{k \in S} 
    \ee^{-\frac{\ii}{m_0} \sin{ \left( 2 \pi k / V \right)}}
    \ee^{-\ii 2 \pi (\ell - \ell') k / V}
\\
\lim_{V \to \infty} \Gamma^\text{01}_{\ell,\ell'} 
&= \frac{1}{2 \pi} \int_0^{2\pi}
    \ee^{\ii(-\frac{1}{m_0} \sin{k} - (\ell - \ell') k)} \, \dd{k}
\\
&= J_{\ell-\ell'} \left( -\frac{1}{m_0} \right),
\end{align}
where $J$ is the Bessel function of the first kind.
Note also that 
$J_{-\alpha}\left(\frac{1}{m_0}\right) 
= (-1)^\alpha J_\alpha\left(\frac{1}{m_0}\right)$.

For $\frac{1}{m_0} << \sqrt{\abs{\ell - \ell'}+1}$,
\begin{align}
J_\alpha \left(\frac{1}{m_0}\right) 
&\approx \frac{1}{\Gamma(\alpha + 1)} \left(\frac{1}{2m_0}\right)^\alpha 
= \frac{1}{\alpha !} \left(\frac{1}{2m_0}\right)^\alpha \nonumber
\\
&\approx \frac{1}{\sqrt{2 \pi \alpha}} 
    \ee^{\alpha (1 - \ln{2 m_0\alpha} )}
\end{align}
And so,
\begin{equation}
\lim_{V \to \infty} \Gamma^\text{01}_{\ell,\ell'} 
\approx \frac{1}{\sqrt{2 \pi (\ell - \ell')}} 
    \ee^{(\ell-\ell') \left( 1 - \ln{2 m_0 (\ell - \ell')} \right) }
\end{equation}

Alternatively, in the limit as the number of modes $V \to \infty$, 
\cref{eqn:fermionic_correlations/correlation_submatrix} becomes
\begin{align}
\lim_{V \to \infty} \Gamma^\text{01}_{\ell,\ell'} &= \frac{1}{2 \pi} 
\int_0^{2\pi} \ee^{-\ii(2\theta_k + (\ell - \ell') k)} \, \dd{k}
\\
\begin{split}
&= \frac{1}{2 \pi} \Bigl(
\int_0^{\pi} \ee^{-\ii(2\theta_k + (\ell - \ell') k)} \, \dd{k}
\\
& \qquad\qquad + \int_0^{\pi} \ee^{\ii(2\theta_k + (\ell - \ell') k)} \, \dd{k}
\Bigr)
\end{split}
\\
&= \frac{1}{\pi} 
\int_0^\pi \cos{(2\theta_k + (\ell - \ell') k)} \, \dd{k},
\end{align}
where, as before, 
$\theta_k = \arctan{(-q_k/(m_0 + \sqrt{m_0^2 + q_k^2}))}$ 
and $q_k$ has been redefined to 
$q_k = 2 \sum_{j=1}^{2\dbK-2} \Delta^{(1)}_j \sin{(j k)}$.
Letting $\theta_k = \arctan{(A_k)}$, 
$A_k = -q_k/(m_0 + \sqrt{m_0^2 + q_k^2})$, 
this can be further simplified to
\begin{gather}
\begin{split}
\lim_{V \to \infty} \Gamma^\text{01}_{\ell,\ell'}
&= -\frac{2}{\pi} \int_0^\pi \frac{A_k}{1+A_k^2} 
    \bigl( A_k \cos{((\ell - \ell') k)} 
\\
& \qquad + \sin{((\ell - \ell') k)} \bigr) \, \dd{k}
\end{split} \\
= \begin{cases}
    -\frac{4}{\pi} 
    \int_0^{\pi/2} B_k \sin{((\ell - \ell') k)} \, \dd{k}, 
    & \ell - \ell'\text{ odd} \\
    -\frac{4}{\pi} 
    \int_0^{\pi/2} C_k \cos{((\ell - \ell') k)} \, \dd{k}, 
    & \ell - \ell'\text{ even}
\end{cases}, \nonumber \\
B_k = \frac{A_k}{1+A_k^2}, \quad C_k = \frac{A_k^2}{1+A_k^2}
\end{gather}

In the special case of Haar wavelets ($\dbK = 1$), 
$\Delta^{(1)}_1 = -\frac{1}{2}$, 
$\Delta^{(1)}_{\ell>1} = 0$, and so 
$q_k = - \sin{ \left( 2 \pi k / V \right)}$.
The above then simplifies to 
\begin{align}
B_k &= \frac{\sin{k}(m_0+\sqrt{m_0^2+\sin^2{k}})}
    {\sin^2{k}+(m_0+\sqrt{m_0^2+\sin^2{k}})^2},
\\
C_k &= \frac{\sin^2{k}}
    {\sin^2{k}+(m_0+\sqrt{m_0^2+\sin^2{k}})^2}.
\end{align}

\end{document}